\documentstyle[12pt,aas2pp4]{article}

\begin{document}
\lefthead{Brogan et al.}
\righthead{Magnetic Fields Toward Five SNR}

\small
% LaTeX definitions.
\newcommand\HII{H\,{\sc ii}}
\newcommand\HI{H\,{\sc i}}
\newcommand\OI{[O\,{\sc i}] 63 $\mu$m}
\newcommand\CII{[C\,{\sc ii}] 158 $\mu$m}
\newcommand\CI{[C\,{\sc i}] 370 $\mu$m}        
\newcommand\SiII{[Si\,{\sc ii}] 35 $\mu$m}
\newcommand\Hi{H110$\alpha$}
\newcommand\He{He110$\alpha$}
\newcommand\Ca{C110$\alpha$}
\newcommand\kms{km~s$^{-1}$}
\newcommand\cmt{cm$^{-2}$}
\newcommand\cc{cm$^{-3}$}
\newcommand\CeO{C$^{18}$O}
\newcommand\tCO{$^{12}$CO}
\newcommand\thCO{$^{13}$CO}
\newcommand\Blos{$B_{los}$}
\newcommand\Bth{$B_{\theta}$}
\newcommand\Bm{$B_{m}$}
\newcommand\Bvm{$\mid\vec{B}\mid$}
\newcommand\BS{$B_S$}
\newcommand\Bscrit{$B_{S,crit}$}
\newcommand\Bw{$B_W$}
\newcommand\mum{$\mu$m}
\newcommand\muG{$\mu$G}
\newcommand\thi{$\tau_{HI}$}
\newcommand\toh{$\tau_{OH67}$}
\newcommand\tmin{$\tau_{min}$}
\newcommand\tmax{$\tau_{max}$}
\newcommand\dv{$\Delta v_{FWHM}$}
\newcommand\va{$v_A$}
\newcommand\Np{$N_p$}
\newcommand\np{$n_p$}
\newcommand\gtsim{\raisebox{-.5ex}{$\;\stackrel{>}{\sim}\;$}}
\newcommand\We{${\cal W}$}
\newcommand\Ms{${\cal M}_S$}
\newcommand\Mw{${\cal M}_w$}
\newcommand\Te{${\cal T}$}
\newcommand\Ps{${\cal P}_s$}
\newcommand\Ra{${\cal R}$}
\newcommand\Da{${\cal D}$}
\newcommand\xb{$x_b$}
\newcommand\xbt{$x_b^2$}

\title{OH Zeeman Magnetic Field Detections Toward Five Supernova Remnants Using the VLA}

\author{C. L. Brogan\altaffilmark{1}, D. A. Frail\altaffilmark{2}, W. M. Goss\altaffilmark{2},
and T. H. Troland\altaffilmark{1}}
\altaffiltext{1}{University of Kentucky, Department of Physics \& Astronomy, 
Lexington, KY 40506-0055}
\altaffiltext{2}{National Radio Astronomy Observatory, P. O. Box O, 1003 Lopezville Road, Socorro, 
NM 87801}

\authoremail{brogan@pa.uky.edu} 

\begin{abstract}
\small

We have observed the OH (1720 MHz) line in five galactic SNRs with the VLA to measure their
magnetic field strengths using the Zeeman effect.  We detected all 12 of the bright ($S_{\nu} >
200$ mJy) OH (1720 MHz) masers previously detected by \markcite{Fra96}Frail et al.  (1996) and
\markcite{Gre}Green et al.  (1997) and measured significant magnetic fields (i.e.  $ >
3\sigma$) in ten of them.  Assuming that the ``thermal'' Zeeman equation can be used to
estimate $\mid\vec{B}\mid$ for OH masers, our estimated fields range from 0.2 to 2 mG.  These
magnetic field strengths are consistent with the hypothesis that ambient molecular cloud
magnetic fields are compressed via the SNR shock to the observed values.  Magnetic fields of
this magnitude exert a considerable influence on the properties of the cloud with the magnetic
pressures ($10^{-7} - 10^{-9}$ erg \cc\/) exceeding the pressure in the ISM or even the thermal
pressure of the hot gas interior to the remnant.  This study brings the number of galactic SNRs
with OH (1720 MHz) Zeeman detections to ten.

\end{abstract}

\keywords\small{ISM:clouds --- ISM:individual (W51, G349.7+0.2, CTB37A, CTB33, G357.7$-$0.1) ---
ISM:magnetic fields --- masers --- polarization --- radio lines:ISM}
\topmargin -0.5in
\textheight9.0in

\section{INTRODUCTION}

Magnetic fields can moderate the impact that a shock has on a molecular cloud.  In the absence
of a field, a supernova blast wave will heat, compress and fragment the cloud and may
ultimately destroy it (\markcite{Kle}Klein, McKee \& Colella 1994).  The inclusion of a field
mitigates these effects, limiting compression and stabilizing it against fragmentation,
allowing the cloud to survive, perhaps to trigger a future generation of star formation
(\markcite{Mie}Miesch \& Zweibel 1994; \markcite{MacL}MacLow et al.  1994).

It has long been known that supernova remnants (SNRs) possess magnetic fields.  Observations of
synchrotron radiation have established that the {\it direction} of the magnetic field in young
SNRs like Cas A is predominately parallel to the shock normal (i.e.  radial), whereas for older
remnants the fields are perpendicular (e.g.  \markcite{Dic}Dickel et al.  1991;
\markcite{Mil}Milne 1990).  The latter geometry likely originates from the compression of the
ambient interstellar field, while Rayleigh-Taylor instabilities are invoked to explain the
radial fields in young SNRs (\markcite{Jun}Jun \& Norman 1996).  Until recently, estimates of
the {\it strength} of the magnetic fields in SNRs had to rely on the somewhat dubious
equipartition approximation.  This situation has changed with the re-discovery of shock-excited
OH (1720 MHz) maser emission in SNRs (\markcite{Fra94}Frail, Goss \& Slysh 1994).

In a series of recent papers, the satellite line of the OH molecule at 1720.53 MHz has been
used as a powerful probe of SNR-molecular cloud interactions.  OH (1720 MHz) masers are found
in $\sim $20 SNRs, or 10\% of the known SNRs in our Galaxy (see \markcite{Kor}Koralesky et al.
1998 and references therein).  This maser line is inverted through collisions with H$_2$
($n\sim$~few$\times{10}^4$ cm$^{-3}$ and T$\sim $80 K) behind C-type shocks propagating into
molecular clouds (\markcite{Rea99}Reach \& Rho 1999, \markcite{Fra98}Frail \& Mitchell 1998).
The geometry of the shock is well constrained since strong maser amplification can only occur
when the shock front is viewed edge on (\markcite{Cla97}Claussen et al.  1997).  These
observational statements are well-supported by theoretical modeling of the OH (1720 MHz)
excitation (\markcite{Loc}Lockett, Gauthier \& Elitzur 1998; \markcite{War99}Wardle,
Yusef-Zadeh, \& Geballe 1999; \markcite{Wardle}Wardle 1999).

One advantage of observing the OH (1720 MHz) maser line is that it allows for a measurement of
the strength of the magnetic field via Zeeman splitting (e.g.  \markcite{Tro}Troland \& Heiles
1982).  In this case, when the observed splitting is small compared to the line width, V $=
ZC$\Bvm\/\thinspace{dI/d$\nu$}, where \Bvm\/ is the the total magnetic field strength, Z is the
Zeeman splitting coefficient, and $C$ is a constant which depends on the angle between the line
of sight and $\vec{B}$ (the possible forms of $C$ will be discussed in \S 4.1).  For now we
will denote the combination of $C$\Bvm\/ as \Bth\/.

This method has already been used to successfully measure the magnetic field strength in the
post-shock gas behind five SNRs (Sgr A East, W44, W28, G32.8$-0.1$, and G346.6$-$0.2), yielding
values for \Bth\/ between 0.1 to 4 mG (\markcite{Yus96}Yusef-Zadeh et al.  1996,
\markcite{Cla97}Claussen et al.  1997, \markcite{Kor}Koralesky et al.  1998).  With these
promising results in mind, we have performed 1720 MHz Zeeman studies toward five more of the 17 SNRs
found in surveys by \markcite{Fra96}Frail et al.  (1996) and \markcite{Gre}Green et al.  (1997)
to contain OH (1720 MHz) masers using the NRAO\footnote{The National Radio Astronomy
Observatory is a facility of the National Science Foundation operated under a cooperative
agreement by Associated Universities, Inc.}  VLA.  Only SNRs with bright masers (I$>$200 mJy)
and no previous high resolution Stokes V observations were chosen.

\section{OBSERVATIONS} 

\placetable{tab1}

\begin{deluxetable}{lcccccrc}
\tablewidth{40pc}
\tablecaption{\footnotesize Observational Parameters with the VLA}
\tablehead{
\colhead{SNR}           & \colhead{R. A.$~^a$}  & \colhead{Decl.$~^a$} & \colhead{$V_{lsr}$} & 
\colhead{$t_{source}$}  &  \colhead{Beam} & \colhead{P. A.} & \colhead{$\sigma_{rms}~^b$} \\
\colhead{}      & \colhead{(B1950)}  & \colhead{(B1950)} & 
\colhead{(km s$^{-1}$)}  &  \colhead{(hours)} & \colhead{($\arcsec$ x $\arcsec$)} & 
\colhead{($\arcdeg$)} & \colhead{(mJy)} }  
\footnotesize\startdata
W51C (2IF) & 19 20 35.0 & +14 09 00.0 & +64.0 & 3.3 & 1.4 x 1.2 & $-$40 & 6.6 \\
~~~~~~~~~~(PA)$^c$ & 19 20 35.0 & +14 09 00.0 & +70.0 & 1.6 & 1.6 x 1.3 & +55 & 6.5 \\
G349.7+0.2 & 17 14 36.0 & $-$37 22 00.0 & +16.0 & 2.6 & 3.4 x 1.3 & +08 & 7.2 \\
CTB37A (I)& 17 11 00.0 & $-$38 30 00.0 & $-$64.0 & 4.9 & 3.9 x 1.2 & $-$14 & 5.4 \\
~~~~~~~~~~~(II) & 17 11 00.0 & $-$38 30 00.0 & $-$22.0 & 4.0 & 4.2 x 2.3 & +23 & 5.5 \\
CTB33 & 16 32 14.0 & $-$47 31 00.0 & $-$70.0 & 2.1 & 8.0 x 1.1 & +67 & 10.2 \\
G357.7$-$0.1 & 17 36 56.0& $-$30 56 00.0 & $-$12.0 & 5.1 & 3.0 x 1.2 & +18 & 5.2\enddata 
\tablenotetext{a} {Units of right ascension are hours, minutes, and seconds, and 
units of declination are degrees, arcminutes, and arcseconds.} 
\tablenotetext{b} {rms noise in an individual channel.}
\tablenotetext{c} {W51C (PA) data have a velocity resolution of 0.56 \kms\/ while all other data have 
velocity resolution of 0.27 \kms\/.}
\label{tab1}
\end{deluxetable}

We observed five SNRs (W51, G349.7+0.2, CTB37A, CTB33, and G357.7$-$0.1) at 1720 MHz with the
VLA in A configuration.  The key observing parameters for each SNR are given in Table 1.  All
of the SNRs were observed in ``2IF mode'' (recording both right (RCP) and left (LCP) circular
polarization) with a 0.1953 MHz (34.0 \kms\/) bandwidth divided into 127 channels.  Due to this
narrow bandwidth, CTB37A had to be observed with two different center frequencies (-22 \kms\/
and -64 \kms\/) in order to obtain data for its two different OH (1720 MHz) maser populations.
The observations were conducted on June 22, July 6, 8, 12, 15, September 13, and October 24,
1999.  The OH data were Hanning smoothed online, and the resulting velocity resolution is 0.27
\kms\/.  We observed both senses of circular polarization simultaneously and since Zeeman
observations are very sensitive to small variations in the bandpass, a front-end transfer
switch was used to periodically switch the sense of circular polarization passing through each
telescope's IF system.

The AIPS (Astronomical Image Processing System) package of the NRAO was used for the
calibration, imaging, and cleaning of the OH (1720 MHz) data sets.  The RCP and LCP data were
calibrated separately and later combined during the imaging process to make Stokes I$=$(RCP +
LCP)/2 and Stokes V$=$(RCP $-$ LCP)/2 data sets.  Bandpass correction was applied only to the I
data sets since bandpass effects cancel to first order in the V data.  Line data sets were
created by estimating and removing the continuum emission in the UV plane using the AIPS task
UVLSF.  The strongest maser channel in each line data set was then self-calibrated and the
solutions were applied to each channel.  Subsequent magnetic field estimates were performed
using the MIRIAD (Multichannel Image Reconstruction Image Analysis and Display) processing
package from BIMA.  The rms noise per spectral channel obtained for each SNR is summarized in
Table 1.

\begin{deluxetable}{lcccrccc}
\tablewidth{40pc}
\tablecaption{\footnotesize Fitted Parameters of OH (1720 MHz)
Masers}
\tablehead{
\colhead{SNR}    &  \colhead{Feature}      & \colhead{R. A.$~^a$}  & \colhead{Decl.$~^a$} & 
\colhead{$S_{peak}~^b$} & \colhead{$v_{lsr}~^c$}  &  \colhead{$\Delta v_{FWHM}~^d$} & 
\colhead{$\theta_{max}$} \\
\colhead{}    &  \colhead{}      & \colhead{(B1950)}  & \colhead{(B1950)} & 
\colhead{(mJy)} & \colhead{(km s$^{-1}$)}  &  \colhead{(km s$^{-1}$)} & 
\colhead{($\arcsec$)} } 
\footnotesize \startdata
W51C ........& 1 & 19 20 35.9 & +14 09 53.6 & 2710 & +72.0 & 0.9 & 0.7 \\
& 2 & 19 20 36.4 & +14 09 50.3 & 4760 & +69.1 & 1.2 & 0.1 \\
G349.7+0.2 ..& 1 & 17 14 36.0 & $-$37 23 01.9 & 770 & +16.2 & 1.1 & 1.6\\
& 2 & 17 14 36.9 & $-$37 22 52.8 & 1800 & +15.2 & 0.3 & 1.0 \\
& 3 & 17 14 37.5 & $-$37 23 14.7 & 1060 & +16.7 & 0.6 & 0.9 \\
CTB37A ......& 1 & 17 10 51.8 & $-$38 28 51.7 & 410 & $-$66.2 & 0.7 & 0.6 \\
& 2 & 17 10 56.8 & $-$38 28 56.7 & 730 &$-$63.7 & 0.5 & 0.3 \\
& 3 & 17 10 59.3 & $-$38 31 19.2 & 470 &$-$63.5 & 1.5 & 0.2 \\
& 4 & 17 11 01.4 & $-$38 37 32.9 & 220 &$-$65.3 & 1.1 & 0.7 \\
& 5 & 17 11 09.0 & $-$38 25 51.3 & 400 &$-$21.5 & 0.9 & 1.7 \\
CTB33 .......& 1 & 16 32 06.2 & $-$47 29 53.2 & 250 & $-$71.8 & 1.0 & 1.7 \\
G357.7$-$0.1 ..& 1 & 17 36 54.6 & $-$30 56 07.6 & 400 & $-$12.3 & 0.6 & 0.8\enddata
\tablenotetext{a} {Units of right ascension are hours, minutes, and seconds, and 
units of declination are degrees, arcminutes, and arcseconds.} 
\tablenotetext{b} {Errors in the peak flux range from 4 to 8 mJy.}
\tablenotetext{c} {Errors in $v_{lsr}$ range from 0.02 to 0.002 \kms\/.}
\tablenotetext{d} {Errors in $\Delta v_{FWHM}$ range from 0.05 to 0.003 \kms\/, and 
the values shown have been deconvolved from the finite channel width (0.27 \kms\/).}
\label{tab2}
\end{deluxetable}

After imaging, each of the bright maser spots was fit with a 2-D Gaussian using the AIPS task
JMFIT.  None of these fits showed convincing evidence that the individual maser spots are
resolved at the resolutions shown in Table 1; therefore, we regard the maximum sizes reported
by JMFIT to be upper limits.  The positions, peak flux densities, and an upper limit to the
maser spot sizes are reported in Table 2.  In addition, the Stokes I spectrum at the peak pixel
of each maser spot was fit with a single Gaussian in the spectral domain using GIPSY, to obtain
each maser's center velocity ($v$) and linewidth ($\Delta v_{FWHM}$).  These estimates of
$\Delta v_{FWHM}$ were then corrected for the finite channel width of the the data (0.27
\kms\/).  These deconvolved $\Delta v_{FWHM}$ and center velocities are also reported in Table
2.  The absolute position errors of these data are $\sim 0.1\arcsec$, while the relative
position errors (compared to other masers in the field) are $\sim 0.02\arcsec$.

In addition, W51 was observed on August 3, 1999 in ``PA mode'' (recording all Stokes
parameters) with a 0.1953 MHz bandwidth, and 64 channels.  This correlator setup resulted in a
velocity resolution of $\sim 0.54$ \kms\/ after Hanning smoothing.  The details of this
observation can also be found in Table 1.  These data were calibrated in the same manner
described above, with the exception of the polarization calibration.  The absolute polarization
calibration was carried out by extrapolating 3C48 data from seven previous 20 cm polarimetry
observations to yield a position angle of P.A.$=13\arcdeg$ for this calibrator at 1720 MHz.
The error in this estimate ($\sim 5\arcdeg$) will also apply to the position angles derived
from these data.

\placetable{tab2}

\section{RESULTS}

\subsection{General Maser Properties}

\begin{deluxetable}{lcc}
\tablewidth{20pc}
\tablecaption{\footnotesize Magnetic Fields}
\tablehead{
\colhead{SNR}    &  \colhead{Feature}      & \colhead{$B_{\theta}~^a$}  \\
\colhead{}    &  \colhead{}      & \colhead{(mG)}  } 
\footnotesize \startdata
W51C ...........& 1 & +1.5 $\pm$ 0.05 \\
& 2 & +1.9 $\pm$ 0.10 \\
G349.7+0.2 ...& 1 & complex~$^b$\\
& 2 & $\lesssim 0.1~^c$ \\
& 3 & $-0.35 ~\pm$ 0.05\\
CTB37A ........& 1 & $-0.5 ~\pm$ 0.10\\
& 2 & +0.22 $\pm$ 0.05\\
& 3 & $-0.60 ~\pm$ 0.09\\
& 4 & +1.5 $\pm$ 0.20\\
& 5 & $-0.8 ~\pm$ 0.10\\
CTB33 .........& 1 & +1.1 $\pm$ 0.30 \\
G357.7$-$0.1 ....& 1 & +0.7 $\pm$ 0.12 \enddata
\tablenotetext{a} {These \Bth\/ field estimates were calculated using the thermal Zeeman equation
and may {\em overestimate} the maser fields according to the maser theory of Elitzur
(1998).  The correction cannot be determined exactly but is probably on the order of $(0.5 -
0.2)\times$\Bth\/, therefore these values should be viewed as upper limits.  See \S 4.1 for further
discussion.}
\tablenotetext{b} {Stokes V profile is complex, probably indicative of blending.}
\tablenotetext{c} {Upper limit to \Bth\/ based on S/N of data. See also \S 3.2.2.}
\label{tab3}
\end{deluxetable}

From the five observed SNRs, we detected all 12 bright OH (1720 MHz) masers (S $>$ 200 mJy)
previously known from the surveys of \markcite{Fra96}Frail et al.  (1996) and
\markcite{Gre}Green et al.  (1997).  Table 2 summarizes the fitted parameters of these maser
features.  Of these, ten show classical S-shaped Stokes V profiles (e.g.
\markcite{Eli98}Elitzur 1998), one shows a complicated Stokes V profile, and one shows no
discernible V signal despite its $\sim 1800$ mJy peak flux density.  The observed masers have
typical deconvolved line widths of $\sim 0.8$ \kms\/, spanning the range from 0.3 to 1.5
\kms\/.  In addition, none of the observed OH (1720 MHz) masers appear to have undergone
significant changes in flux density or position since the detection experiments were performed
in 1994.  As noted in \S 2, none of the maser spots have been resolved at the resolutions
listed in Table 1.  This may be due, in part, to the low declination, and hence elliptical
synthesized beam shapes observed for four of the five SNRs.  Using the upper limits on the
maser spot sizes listed in Table 2, lower limits on the brightness temperatures of these masers
range from $4 \times 10^4$ K to $10^8$ K. 

Using the thermal Zeeman expression (V $=ZC$\Bvm\/\thinspace{$d$I$/d\nu$}) with $Z=0.6536$ Hz
\muG\/$^{-1}$ for OH at 1720 MHz, the derivative of Stokes I was fitted to Stokes V in a least
square fitting routine as described by \markcite{Cla97}Claussen et al (1997), to obtain the
combination $C$\Bvm\/=\Bth\/.  The validity of the thermal Zeeman formulation for the magnetic field
in masers along with the $\theta$ dependence and magnitude of $C$ is discussed in \S 4.1.  Magnetic
fields were detected toward all five SNR at greater than the 3$\sigma$ level.  The fitted magnetic
field strengths (\Bth\/), and their associated $1\sigma$ errors are summarized in Table 3, while
comments on individual sources appear in \S 3.2.  The detected \Bth\/ fields range from 0.2 to 2 mG.
Unlike W28 and W44, where the fields were found to be uniform in both magnitude and direction
(\markcite{Cla97}Claussen et al.  1997), CTB37A shows a complicated magnetic field morphology.  In
this source \Bth\/ changes by a factor of seven and reverses sign on a length scale of $\sim 3$ pc.

\placetable{tab3}

\subsection{Individual Sources}

\subsubsection{W51}

\begin{figure}[ht]
\epsfxsize=8.5cm \epsfbox{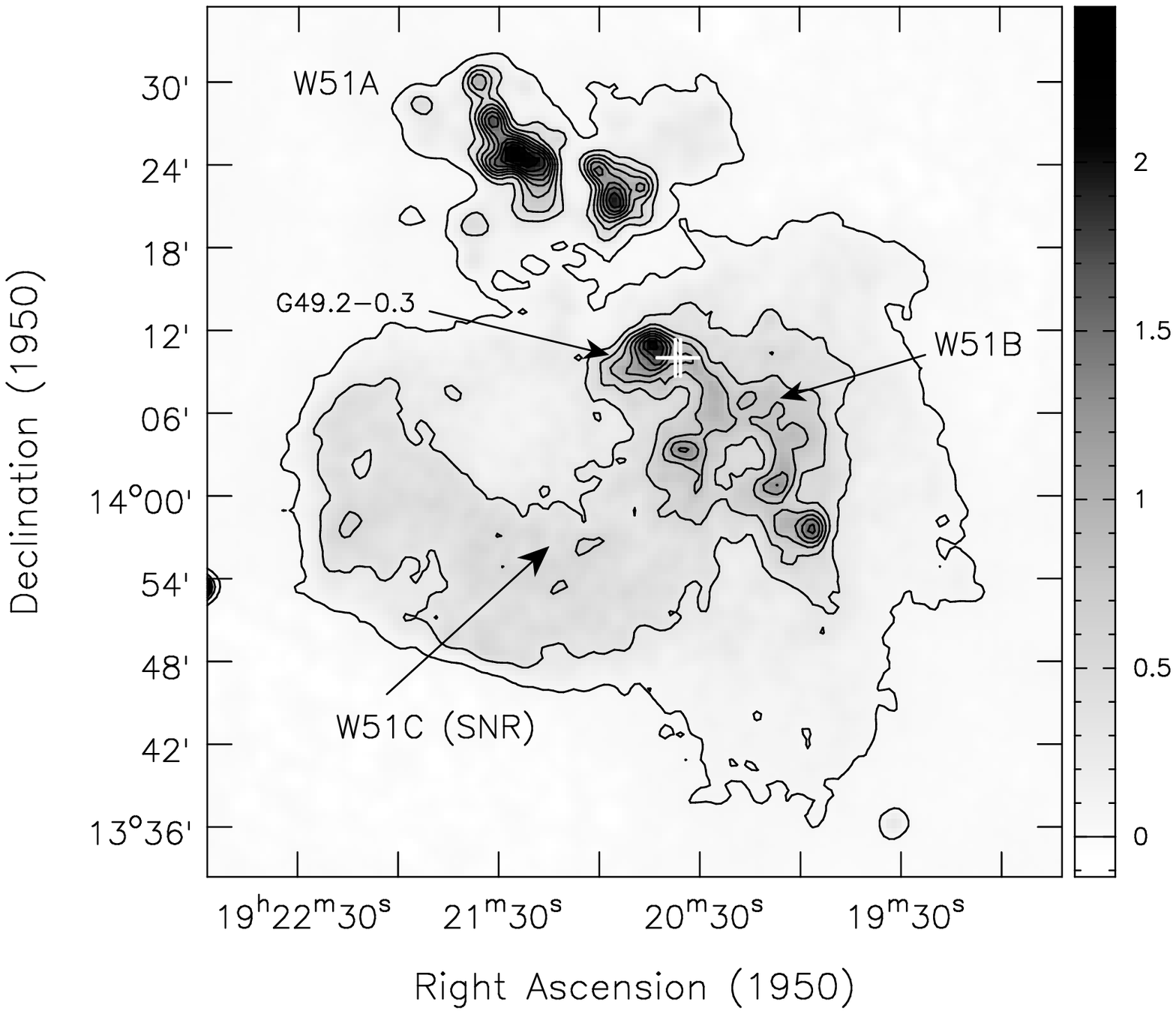}
\figcaption[Filename]{\footnotesize W51 VLA 330 MHz continuum image from Subrahmanyan \& Goss (1995) with
contours at (0.02, 0.12, 0.22, 0.32, 0.42, 0.52, 0.62, 0.72, 0.82) $\times$ 2.46 Jy beam$^{-1}$
.  The resolution of this image is $\sim 1\arcmin$.  The locations of the two W51C OH (1720
MHz) masers are indicated by the white (+) symbols\label{figu1}.}
\end{figure}

\placefigure{figu1}

W51 is composed of two \HII\/ region complexes (W51A and W51B), as well as the SNR W51C and is
located at the tangent point of the Sagittarius arm at $\ell=49\arcdeg$ corresponding to a
distance of $\sim 6$ kpc (\markcite{Koo95}Koo, Kim, \& Seward 1995).  A continuum image of W51
at 330 MHz from \markcite{Sub}Subrahmanyan \& Goss (1995) is shown in Figure 1 with the two W51
OH (1720 MHz) masers superposed.  The spectral index of W51C is difficult to calculate due to
contamination from the W51B \HII\/ regions, but \markcite{Sub}Subrahmanyan \& Goss (1995)
estimate that $\alpha = -0.2$ ($S=\nu^{\alpha}$).

\begin{figure}[ht]

\epsfxsize=9.0cm \epsfbox{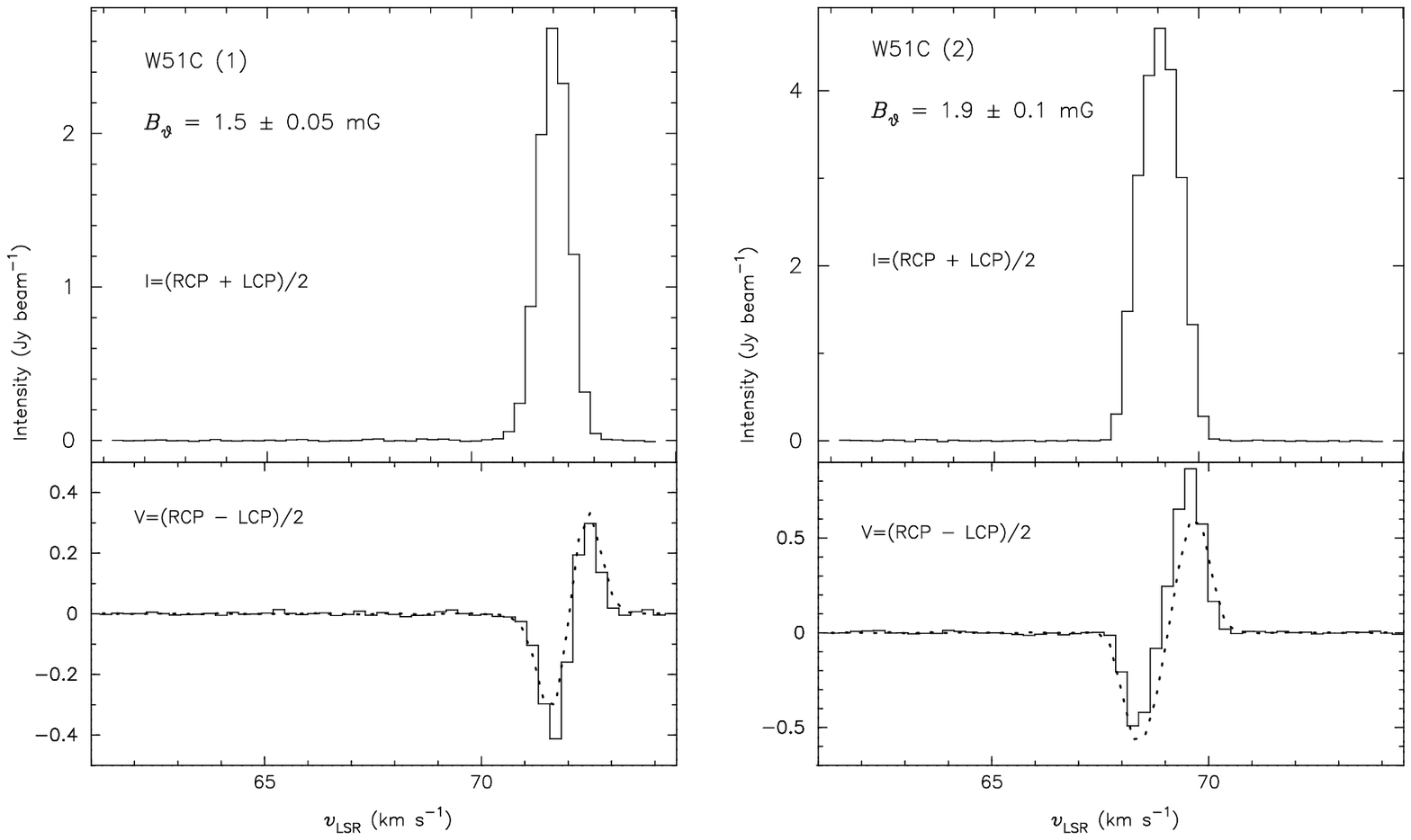}
\figcaption[Filename]{\footnotesize Fits of \Bth\/ for W51 OH (1720 MHz) maser features (1) and (2).  The upper
panels show the VLA Stokes I profiles ({\em solid histogram}), and the bottom panels show the
VLA Stokes V profiles ({\em solid histogram}) with the fitted derivative of Stokes I shown as
smooth dotted curves.  The value of \Bth\/ fit for each position and its calculated error are
given at the top of each plot\label{figu2}.}
\end{figure}

Based on the absorption of X-ray emission from W51C toward W51B, \markcite{Koo95}Koo et al.
(1995) suggest that W51C must be located behind W51B.  \markcite{Koo97a}\markcite{Koo97b}Koo \&
Moon (1997a, 1997b) observed high velocity (HV) \HI\/, CO(1$-$0), and CO(2$-$1) emission
between $+85$ and $+120$ \kms\/ to the east of W51B where it overlaps W51C (see Fig.  1).  They
interpret this HV gas as arising from a shock interaction between W51C and a molecular cloud
(located between W51C and W51B, or possibly the backside of the W51B complex itself).  They
show that the HV \HI\/ is located toward the western edge of the centrally bright X-ray
emitting region of W51C in an arc-shape, while the shocked CO gas is located slightly east of
the HV \HI\/ emission.  The two OH (1720 MHz) masers reported in this work (and
\markcite{Gre}Green et al.  1997) are located toward the NE end of the shocked CO/\HI\/ arc
structure and $\sim 2\arcmin$ SW of the \HII\/ region G49.2-0.3 (see Fig.  1) at velocities of
$\sim 70$ \kms\/.  From the coincidence of these OH (1720 MHz) masers with the shocked gas and
their angular separation from G49.2$-$0.3 ($\sim 3.5$ pc for $d=6$ kpc), \markcite{Gre}Green et
al.  (1997) argue that they are associated with the W51C SNR shock.

\begin{figure}[ht]

\epsfxsize=5.0cm \epsfbox{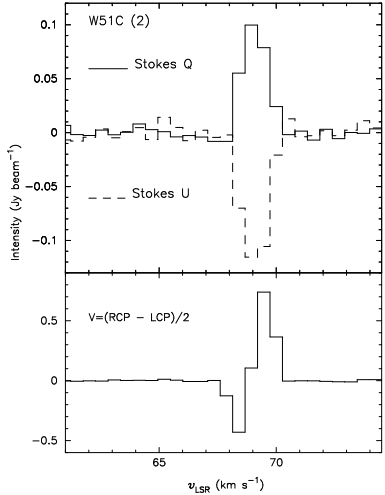}
\figcaption[Filename]{\footnotesize VLA Stokes Q and U linear polarization profiles ({\em upper panel}) and
Stokes V circular polarization profile ({\em lower panel}) for W51C OH (1720 MHz) maser feature
(2). Note that the spectral resolution of these data are only 0.56 \kms\/ 
while the spectra displayed in Fig. 2 (and all other Figures) have a resolution of 0.27 \kms\/\label{figu3}.}
\end{figure}

\placefigure{figu2}
\placefigure{figu3}

Fits of \Bth\/ for the two W51C OH (1720 MHz) masers are shown in Figure 2.  The values of
\Bth\/ for these masers are 1.5 and 1.9 mG for features 1 and 2, respectively (see Tables 2 and
3).  The W51C masers were also observed in PA mode, providing images of all four Stokes
parameters (I, V, Q, U), with 0.54 \kms\/ velocity resolution.  Toward the strongest W51C OH
maser (feature 2), Stokes Q and U signals were detected at 15$\sigma$ and 17$\sigma$,
respectively.  A profile showing Stokes Q and U toward this maser is shown in Figure 3.  Using
Stokes Q and U values of Q $= 0.10$ Jy beam$^{-1}$ and U $= -0.11$ Jy beam$^{-1}$, we obtain a
linear polarization percentage of 3.5\% and a P.A.  of $-25^{\circ}$ for the linear polarization 
vector.

\subsubsection{G349.7+0.2}

\begin{figure}[ht] 
\epsfxsize=8.5cm \epsfbox{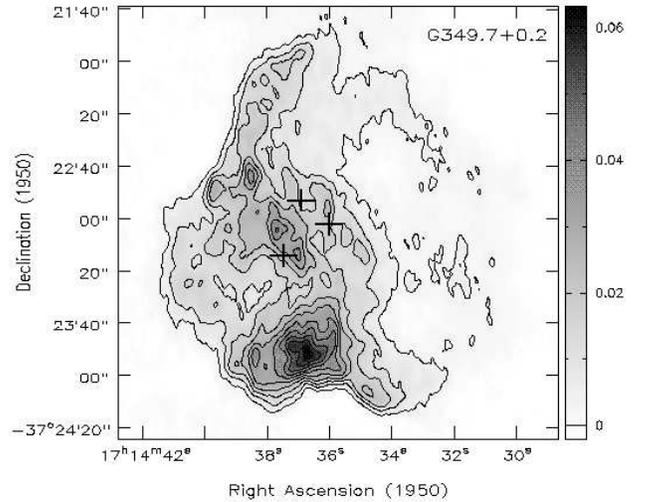}
\figcaption[Filename]{\footnotesize G349.7+0.2 20 cm VLA continuum image with contours at (0.05, 0.15, 0.25,
0.35, 0.45, 0.55, 0.65, 0.75, 0.85, 0.95) $\times$ 63 mJy beam$^{-1}$.  This image is composed
of archival VLA data from A, B, C, and D configurations.  The resolution is $5\arcsec \times
2\arcsec$ (P.A.  = 0.6\arcdeg) and the peak flux density is 63 mJy beam$^{-1}$ with an rms
noise of $\sim 1$ mJy beam$^{-1}$.  The locations of the three G349.7+0.2 OH (1720 MHz) masers
with $S_{\nu}> 200$ mJy are indicated by the black (+) symbols\label{figu4}.}  
\end{figure}

\placefigure{figu4}

\begin{figure}[ht]

\epsfxsize=5.0cm \epsfbox{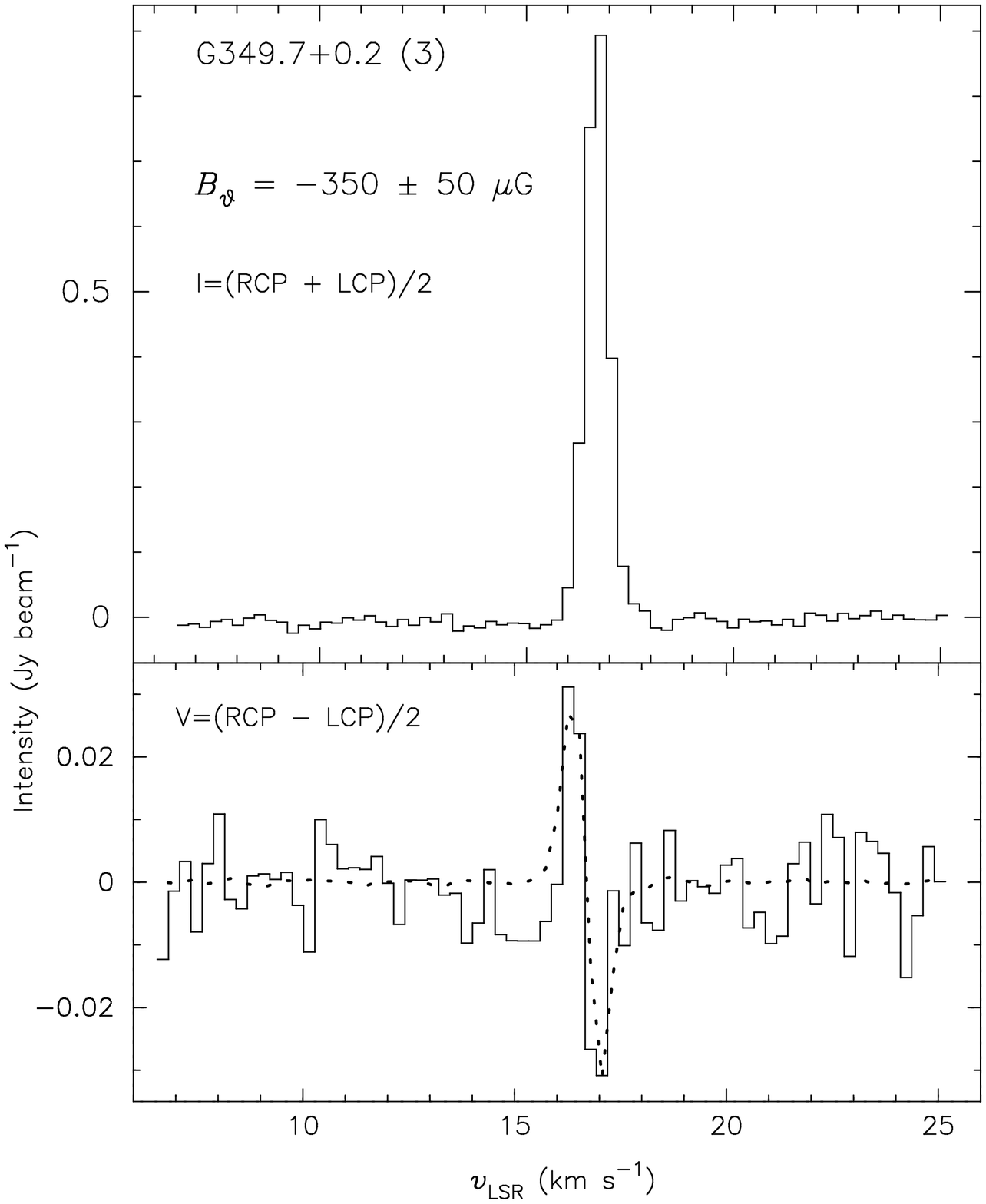} 
\figcaption[Filename]{\footnotesize Fit of \Bth\/ for G349.7+0.2 OH (1720 MHz) maser feature (3).  The
upper panels show the VLA Stokes I profiles ({\em solid histogram}), and the bottom panels show the
VLA Stokes V profiles ({\em solid histogram}) with the fitted derivative of Stokes I shown as smooth
dotted curves.  The value of \Bth\/ fit for each position and its calculated error are given at the
top of each plot\label{figu5}.}
\end{figure}

G349.7+0.2 is one of the most luminous SNR's in the galaxy (after Cas A and the Crab;
\markcite{Shaa}Shaver et al.  1985a), if it is located at a distance of $\sim 22$ kpc
(\markcite{Fra96}Frail et al.  1996).  The spectral index of G349.7+0.2 was estimated by
\markcite{Shaa}Shaver et al.  (1985a) to be $\sim -0.5$, typical of shell type remnants.  This
SNR contains three bright maser features along with several weaker features within $\sim
1\arcmin$ (6 pc) near the center of the remnant.  The positions of the brightest masers are
shown on a continuum image of G349.7+0.2 in Figure 4.  This continuum image is a compilation of
18 and 20 cm data from the VLA archive, and contains data from all four configurations (A, B,
C, and D).  It is the most sensitive and highest resolution continuum image of this SNR to
date, with a resolution of $5\arcsec \times 2\arcsec$ (P.A.$=0.6\arcdeg$) and rms noise of
$\sim 1$ mJy beam$^{-1}$.

\placefigure{figu5}

Only G349.7+0.2 OH(3) (OH maser feature 3) has a significant magnetic field detection with \Bth\/$ =
0.35$ mG.  The \Bth\/ fit for this feature is shown in Figure 5.  Maser features 1 and 2 are the
only bright (S $>$ 200 mJy) OH (1720 MHz) masers in our sample for which we were unable to detect a
significant \Bth\/.  G349.7+0.2 OH(1) exhibits a complicated Stokes V spectrum that is indicative of
blending.  Therefore, it is possible that improved velocity and/or angular resolution would lead to
a \Bth\/ detection for this maser.  G349.7+0.2 OH(2) shows no hint of any Stokes V signal despite
its 1800 mJy flux density.  The S/N of these data alone suggest that \Bth\/$ < 0.1$ mG, however,
this maser is also the narrowest maser in our sample ($\Delta v_{FWHM}=0.3$ \kms\/).  Therefore, it
is possible that if the Stokes V was not resolved in velocity, \Bth\/ for 
G349.7+0.2 OH(2) could be $\sim 0.3$ mG.

\subsubsection{CTB37A}

CTB37A, also known as G348.5+0.1, is estimated to lie at a distance of $\sim 11$ kpc
(\markcite{Fra96}Frail et al.  1996).  \markcite{Kas}Kassim, Baum, \& Weiler (1991) estimate
its spectral index to be $\sim -0.5$ based on flux density measurements ranging from 80 MHz to
14.7 GHz.  A 21 cm continuum map of CTB37A from \markcite{Kas}Kassim et al.  (1991) is
displayed in Figure 6 with the OH (1720 MHz) masers superposed.  CTB37A contains two
kinematically distinct sets of OH (1720 MHz) masers.  One group at $\sim -22$ \kms\/ is located
toward the north end of CTB37A, while the others have velocities of $\sim -65$ \kms\/ and are
located near the center and southern parts of the source.  \markcite{Kas}Kassim et al.  (1991)
propose that the extension of continuum emission seen to the east of CTB37A (see Fig.  6) is a
separate SNR which they name G348.5$-$0.0.  It was further proposed by \markcite{Fra96}Frail et
al.  (1996) that the $\sim -22$ \kms\/ masers originate from this second SNR.

\begin{figure}[t]
\epsfxsize=8.5cm \epsfbox{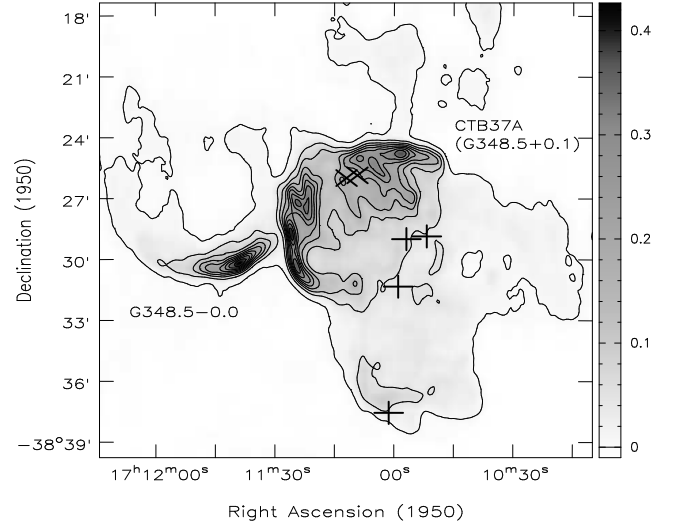 }
\figcaption[Filename]{\footnotesize CTB37A 20 cm VLA continuum image from Kassim et al.  (1991) with contours
at (0.01, 0.12, 0.23, 0.34, 0.45, 0.56, 0.66, 0.76 0.87) $\times$ 0.427 Jy beam$^{-1}$.  The
beam is $33\arcsec \times 18\arcsec$ (P.A.$=5\arcdeg$).  The (+) symbols mark the locations of
the $\sim -65$ \kms\/ OH (1720 MHz) maser features, while the locations of the $\sim -22$
\kms\/ OH (1720 MHz) maser features are marked with ($\times$) symbols\label{figu6}.}
\end{figure}

\placefigure{figu6}

CTB37A, also known as G348.5+0.1, is estimated to lie at a distance of $\sim 11$ kpc
(\markcite{Fra96}Frail et al.  1996).  \markcite{Kas}Kassim, Baum, \& Weiler (1991) estimate
its spectral index to be $\sim -0.5$ based on flux density measurements ranging from 80 MHz to
14.7 GHz.  A 21 cm continuum map of CTB37A from \markcite{Kas}Kassim et al.  (1991) is
displayed in Figure 6 with the OH (1720 MHz) masers superposed.  CTB37A contains two
kinematically distinct sets of OH (1720 MHz) masers.  One group at $\sim -22$ \kms\/ is located
toward the north end of CTB37A, while the others have velocities of $\sim -65$ \kms\/ and are
located near the center and southern parts of the source.  \markcite{Kas}Kassim et al.  (1991)
propose that the extension of continuum emission seen to the east of CTB37A (see Fig.  6) is a
separate SNR which they name G348.5$-$0.0.  It was further proposed by \markcite{Fra96}Frail et
al.  (1996) that the $\sim -22$ \kms\/ masers originate from this second SNR.

\begin{figure}[ht]
\epsfxsize=8.5cm \epsfbox{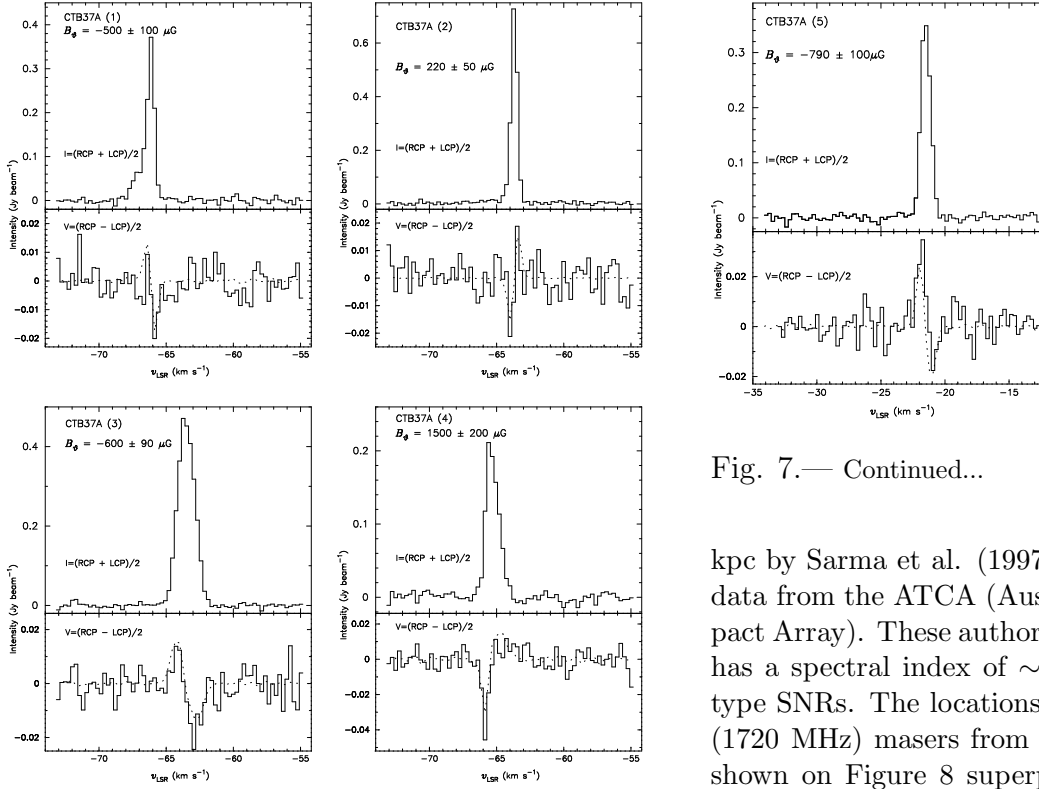}
\figcaption[Filename]{\footnotesize Fits of \Bth\/ for CTB37A OH (1720 MHz) maser features (1 - 5).  Note that
features (1-4) correspond to the $\sim -65$ \kms\/ masers, while feature (5) is the single strong $\sim
-22$ \kms\/ maser.  The upper panels show the VLA Stokes I profiles ({\em solid histogram}), and the
bottom panels show the VLA Stokes V profiles ({\em solid histogram}) with the fitted derivative of
Stokes I shown as smooth dotted curves.  The value of \Bth\/ fit for each position and its
calculated error are given at the top of each plot\label{figu7}.}
\end{figure}

%\clearpage

\addtocounter{figure}{-1}
\begin{figure}[ht]

\epsfxsize=4.5cm \epsfbox{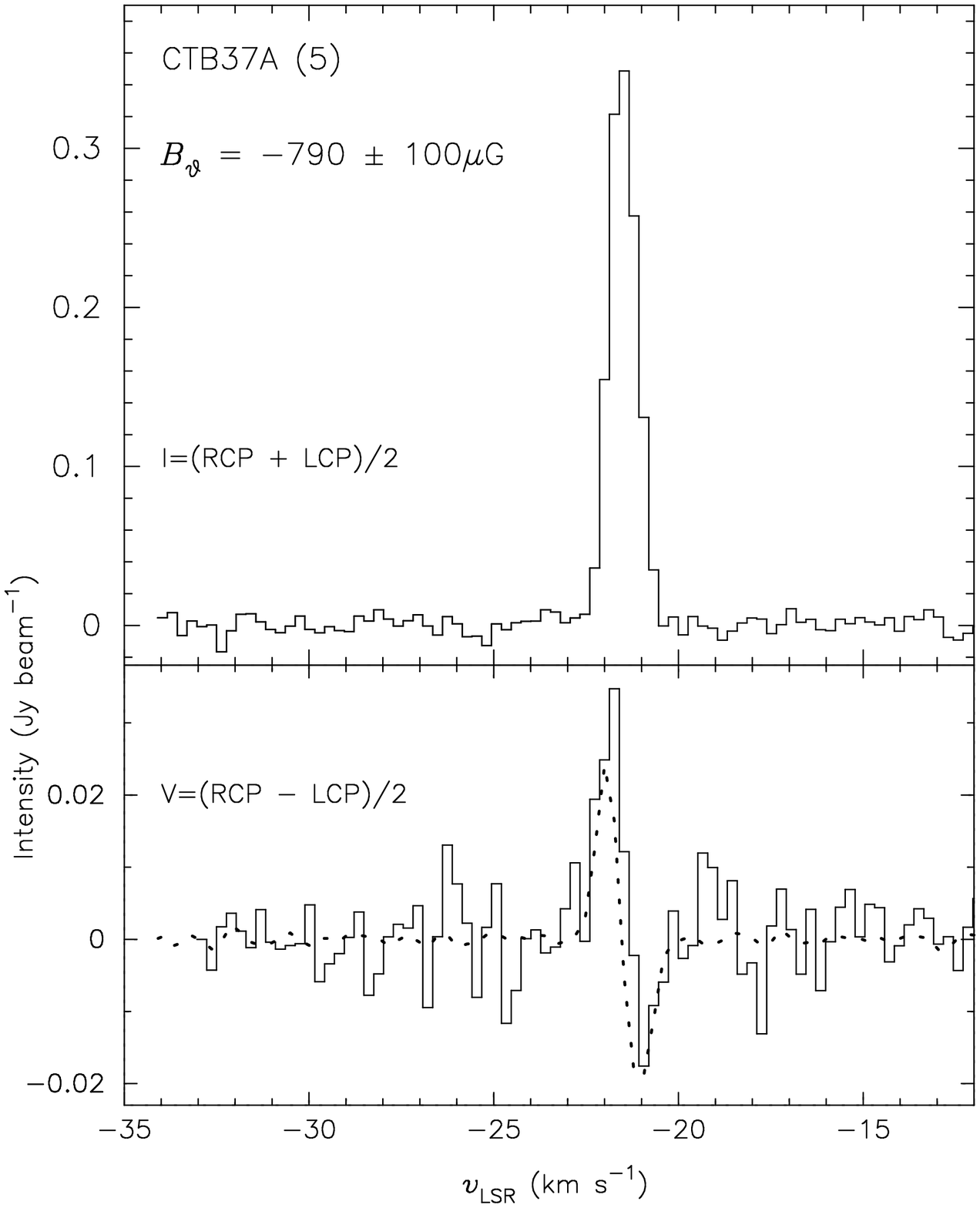}
\figcaption[Filename]{\footnotesize Continued...}
\end{figure}

As noted at the beginning of this section, the \Bth\/ of the $\sim -65$ \kms\/ OH (1720 MHz)
masers have the unusual property that they change direction along the line of site over length
scales as small as $1\arcmin$ ($\sim 3$ pc).  In addition, the magnitude of \Bth\/ changes by a
factor of seven between CTB37A OH(2) and CTB37A OH(4) from $\sim 0.2$ mG to $\sim 1.5$ mG.
Fits of \Bth\/ for these four $\sim -65$ \kms\/ masers are shown in Figure 7.  The only strong
$\sim -22$ \kms\/ OH maser has \Bth\/$= -0.8$ mG and the fit for this maser feature (CTB37A
OH(5)) is also shown in Fig.  7.

Reynoso \& Mangum (1999, in prep.)  have detected CO(1-0) emission toward CTB37A with $\sim
1\arcmin$ resolution using the Kitt Peak 12m telescope.  Their maps show two distinct CO
clouds, one at $\sim -22$ \kms\/ in the northern part of CTB37A, and another at $\sim -65$
\kms\/ which is concentrated to the northwest and middle of the source.  Both CO clouds are
coincident spatially and in velocity with our two groups of OH (1720 MHz) masers.

\placefigure{figu7}

As noted at the beginning of this section, the \Bth\/ of the $\sim -65$ \kms\/ OH (1720 MHz)
masers have the unusual property that they change direction along the line of site over length
scales as small as $1\arcmin$ ($\sim 3$ pc).  In addition, the magnitude of \Bth\/ changes by a
factor of seven between CTB37A OH(2) and CTB37A OH(4) from $\sim 0.2$ mG to $\sim 1.5$ mG.
Fits of \Bth\/ for these four $\sim -65$ \kms\/ masers are shown in Figure 7.  The only strong
$\sim -22$ \kms\/ OH maser has \Bth\/$= -0.8$ mG and the fit for this maser feature (CTB37A
OH(5)) is also shown in Fig.  7.

Reynoso \& Mangum (1999, in prep.)  have detected CO(1-0) emission toward CTB37A with $\sim
1\arcmin$ resolution using the Kitt Peak 12m telescope.  Their maps show two distinct CO
clouds, one at $\sim -22$ \kms\/ in the northern part of CTB37A, and another at $\sim -65$
\kms\/ which is concentrated to the northwest and middle of the source.  Both CO clouds are
coincident spatially and in velocity with our two groups of OH (1720 MHz) masers.

\subsubsection{CTB33}

\begin{figure}[ht]
\epsfxsize=8.5cm \epsfbox{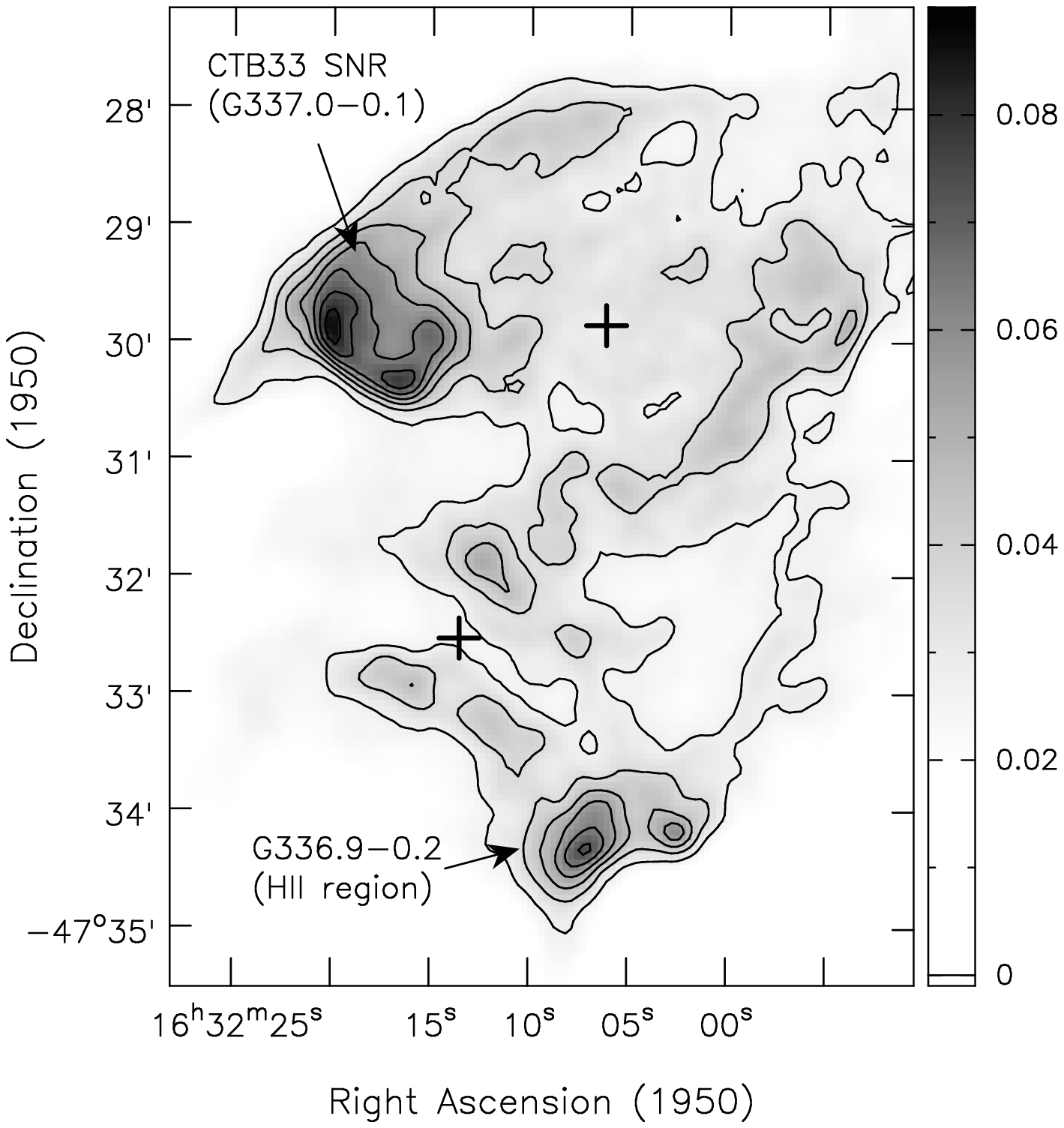}
\figcaption[Filename]{\footnotesize CTB33 1375 MHz ATCA continuum image from Sarma et al.  (1997) with
contours at (0.3, 0.4, 0.5, 0.6, 0.7, 0.8, 0.9) $\times$ 0.090 Jy beam$^{-1}$.  The resolution
of this image is $\sim 12\arcsec$.  The locations of the two CTB33 OH (1720 MHz) masers with
$S_{\nu}> 200$ mJy are indicated by the black (+) symbols\label{figu8}.}
\end{figure}

\placefigure{figu8}

The SNR in CTB33 is also known as G337.0$-$0.1, and was estimated to lie at a distance of $\sim 11$
kpc by \markcite{Sarma}Sarma et al.  (1997) using \HI\/ absorption data from the ATCA (Australia
Telescope Compact Array).  These authors also find that CTB33 has a spectral index of $\sim -0.6$,
typical of shell type SNRs.  The locations of the two bright OH (1720 MHz) masers from
\markcite{Fra96}Frail et al.  (1996) are shown on Figure 8 superposed on a 1380 MHz CTB33 continuum
image (\markcite{Sarma}Sarma et al.  1997).  However, \markcite{Sarma}Sarma et al.  (1997) present
evidence that the southern most CTB33 maser is probably associated with the \HII\/ region
G336.9$-$0.2.  Our OH (1720 MHz) results are consistent with this suggestion since this maser
feature has been resolved into at least four different velocity components, each of which show
complicated Stokes V spectra typical of \HII\/ region masers (i.e.  no ``S'' shaped Zeeman pattern;
see \markcite{Eli98}Elitzur 1998).  For this reason only the centrally located CTB33 maser (CTB33
OH(1)) has been included in Tables 2 and 3.  The \Bth\/ detected for this maser is 1.1 mG, but the
range of channels over which the fit was performed had to be limited to obtain a reasonable fit (see
Fig.  9).  Therefore, this detection should be considered tentative.  Such a suppression of part of
the Stokes V signal may be the result of blending of the maser line in velocity space since this
maser is fairly wide (\dv\/$\sim 1$ \kms\/, also see \markcite{Rob}Roberts et al.  1993), but could also
simply be a result of poor S/N.

\begin{figure}[ht]

\epsfxsize=5.0cm \epsfbox{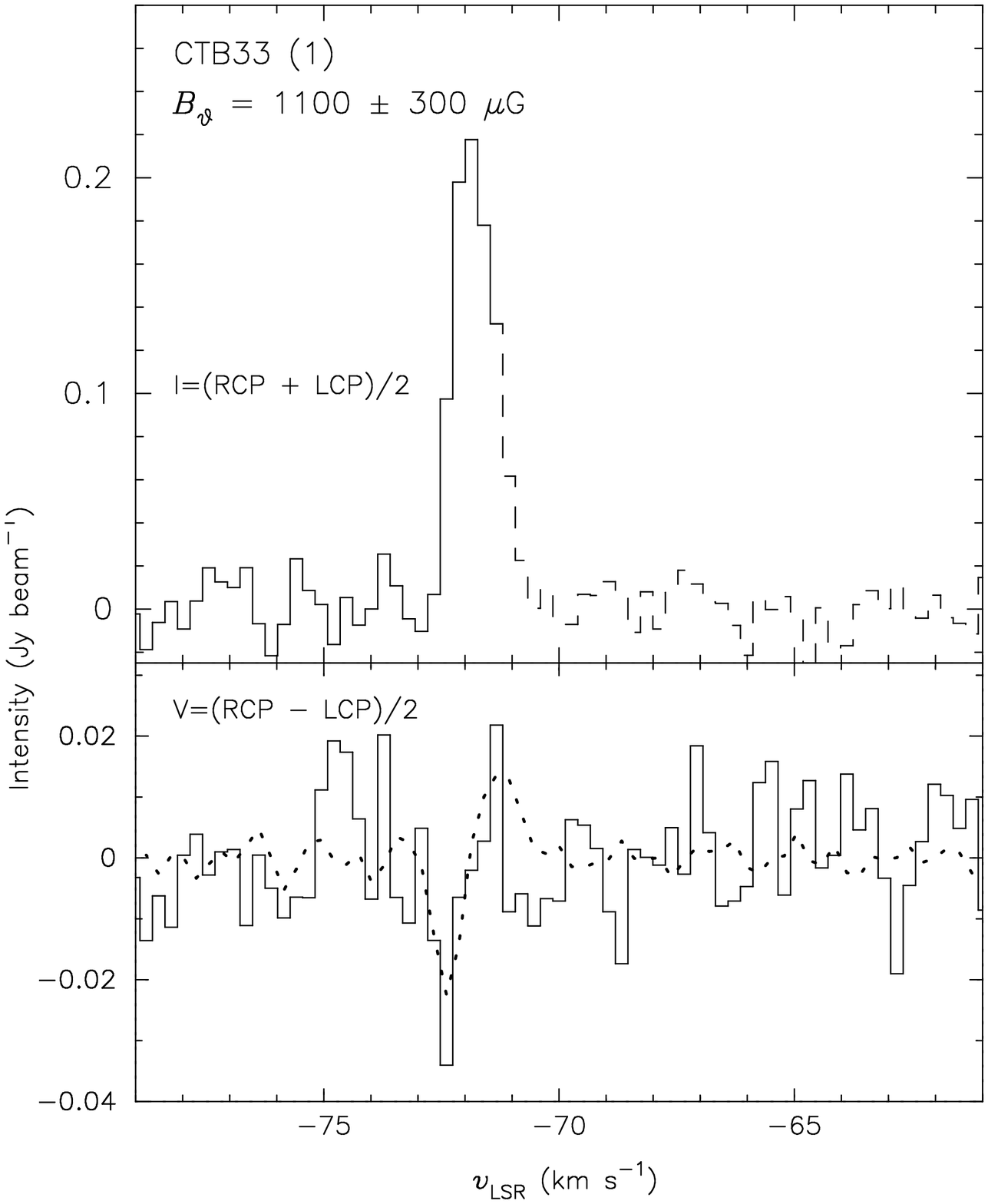}
\figcaption[Filename]{\footnotesize Fit of \Bth\/ for CTB33 OH (1720 MHz) maser feature (1).  The upper panels
show the VLA Stokes I profiles ({\em solid histogram}), and the bottom panels show the VLA Stokes V
profiles ({\em solid histogram}) with the fitted derivative of Stokes I shown as smooth dotted
curves.  The value of \Bth\/ fit for each position and its calculated error are given at the top of
each plot.  The solid portion of the Stokes I histogram ({\em upper panel}) shows the velocity
range used in the fit\label{figu9}.}
\end{figure}

\placefigure{figu9}

\markcite{Cor}Corbel et al.  (1999) have detected a number of molecular clouds toward CTB33 in
CO($J=1-0$) and CO($J=2-1$) emission using the SEST telescope.  In particular, they find a
giant molecular cloud at a velocity of $\sim -71$ \kms\/ ($\Delta v=11$ \kms\/) with an
approximate size of 67 pc and a mass of $4\times 10^6$ M$_{\sun}$.  The velocity of this
molecular cloud is in excellent agreement with the velocity of CTB33 OH(1) ($-71.8$ \kms\/) and
may well be the origin of the SNR/molecular cloud interaction needed to pump the maser (see \S
1 \& \S 4.2).  These authors also suggest that the SNR must lie on the near-side of the $\sim
-71$ \kms\/ GMC based on comparison of their CO data (and its implied extinction) with the
X-ray data from \markcite{Woo}Woods et al.  (1999).

\begin{figure}[ht]
\epsfxsize=8.5cm \epsfbox{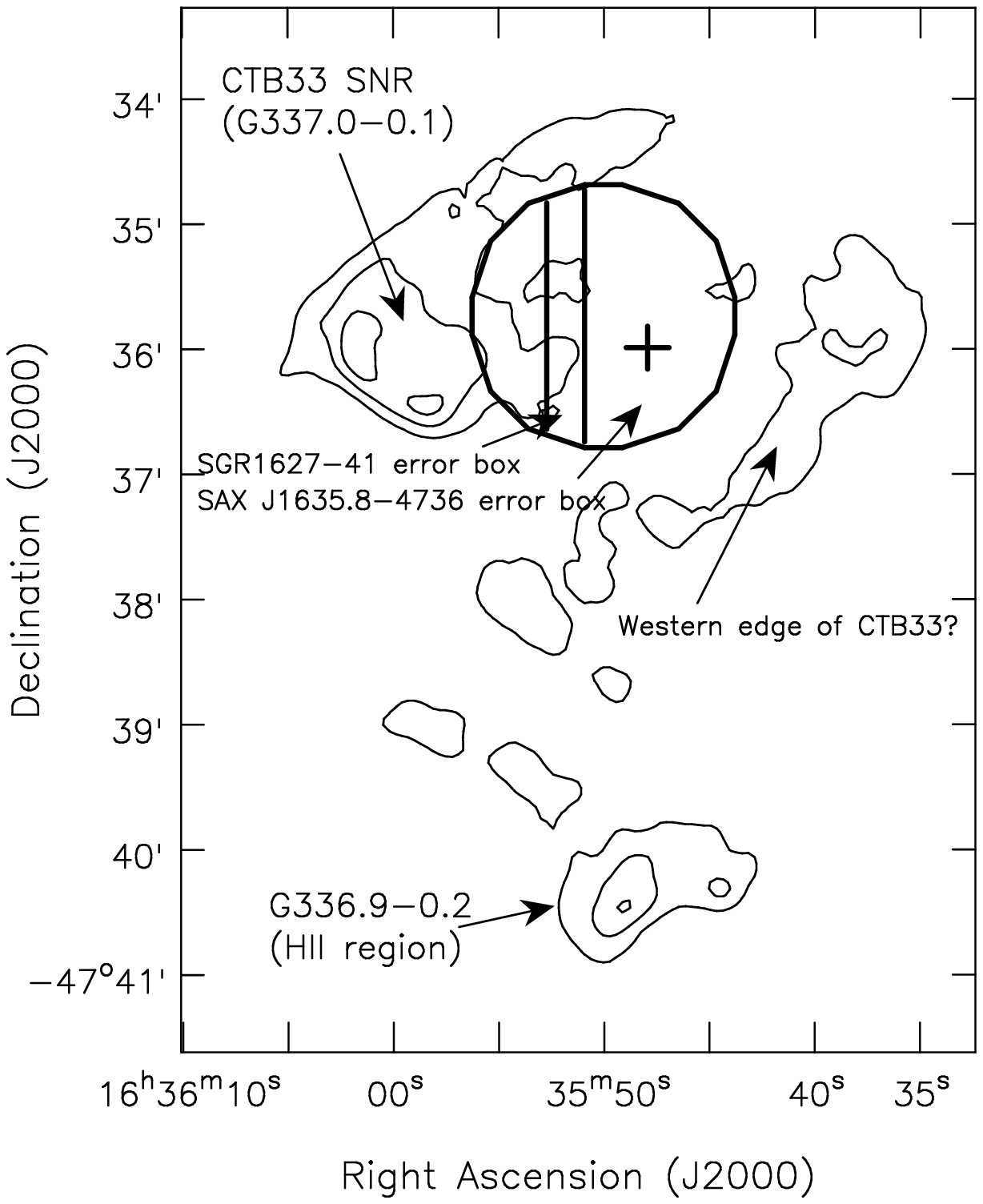}
\figcaption[Filename]{\footnotesize Similar to Figure 8, but in J2000 coordinates with contours at (0.4, 0.6,
0.8) $\times$ 0.090 Jy beam$^{-1}$.  The IPN and BeppoSAX error boxes are superposed showing
the location of SGR1627$-$41\label{figu10}.}
\end{figure}

\placefigure{figu10}

The CTB33 SNR is coincident with the site of a recently discovered Soft Gamma ray Repeater
(SGR1627$-$41; see \markcite{Hur}Hurley et al.  1999).  The location of the SGR has been
constrained by the 3rd Interplanetary Network (IPN:  {\em Ulysses}, KONUS-WIND and BATSE) along
with the detection of a presumably related BeppoSAX X-ray source to lie within the error boxes
shown in Figure 10 on a simplified contour image of CTB33 (\markcite{Sarma}Sarma et al.  1997;
\markcite{Hur}Hurley et al.  1999; \markcite{Woo}Woods et al.  1999).  Current theories for the
nature of SGR's suggest that they arise from strongly magnetized neutron stars or `magnetars',
and that the outbursts are the result of crustquakes on the surface of the neutron star
(\markcite{Tho}Thompson \& Duncan 1995).  The discovery of SGR1627$-$41, marks only the fourth
such source to be detected.

All of the previously known SGRs have been associated with young SNRs (although SGR1900+14 lies
close to, but not inside its associated SNR; see \markcite{Hur}Hurley et al.  1999 and
references therein).  For this reason, despite the inexact correspondence between the
IPN/BeppoSAX error boxes and the extent of CTB33 ($95\arcsec$) proposed by
\markcite{Sarma}Sarma et al.  (1997), SGR1627$-$41 has been assumed to be the progenitor of
CTB33.  For this reason, a number of authors (\markcite{Hur}Hurley et al.  1999;
\markcite{Cor}Corbel et al.  1999) have estimated that SGR1627$-$41 must have a transverse
velocity of more than $\sim 1,000$ \kms\/ if CTB33's age is $\lesssim 5,000$ years.  Although
these velocities are not unreasonable (\markcite{Lyn}Lyne \& Lorimer 1994), the location of the
maser CTB33 OH(1) suggests that CTB33 actually extends farther west as shown in Figure 10, and
has a `blowout' morphology toward the NE and SW.  Such a morphology would not be unexpected in
a region populated by \HII\/ regions that could have effectively cleared such cavities (see
\markcite{Jon}Jones et al.  1998).  Therefore, an increase in the estimated size of CTB33 would
place the IPN/BeppoSAX error boxes and the OH (1720 MHz) maser much closer to the center of
CTB33, rather than at its outskirts, with a substantial reduction in the estimate for the
transverse velocity of SGR1627$-$41.  Future molecular, or high resolution, low-frequency
continuum data toward CTB33 may help resolve this issue.

\subsubsection{G357.7$-$0.1 (Tornado)}

\begin{figure}[ht]
\epsfxsize=9.0cm \epsfbox{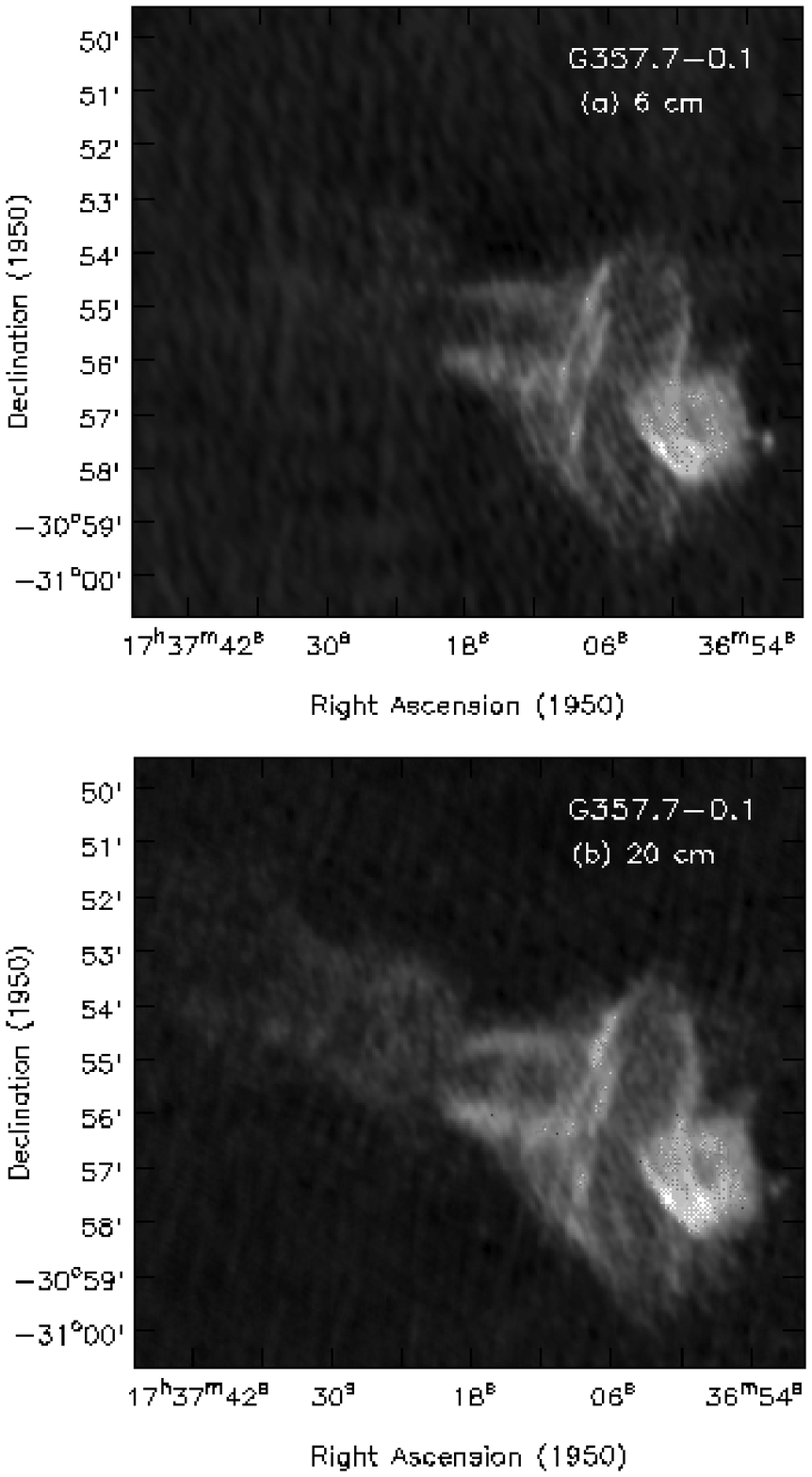}
\figcaption[Filename]{\footnotesize Greyscale continuum images of G357.7$-$0.1 composed of archival 
VLA data showing its spiral morphology.  a) 6 cm image with a resolution of $13\arcsec.4 \times
6\arcsec.7$ (P.A.  = 23.8\arcdeg) and a peak flux density of 77 mJy beam$^{-1}$ with an rms
noise of $\sim 0.6$ mJy beam$^{-1}$.  b) Same as a) but at 20 cm.  The resolution of the 20 cm
image is $11\arcsec.6 \times 8\arcsec.1$ (P.A.  = 26.2\arcdeg) and the peak flux is 145 mJy
beam$^{-1}$ with an rms noise of $\sim 1.5$ mJy beam$^{-1}$\label{figu11}.}
\end{figure}

\placefigure{figu11}

G357.7$-$0.1 is an unusual SNR candidate located near the galactic center with a non-thermal
spectral index in the range $-0.5 < \alpha < -1.0$ (\markcite{Ste}Stewart et al.  1994).  It
has been variously considered to be everything from an extragalactic head-tail or double lobed
source (\markcite{Wei}Weiler \& Panagia 1980; \markcite{Cas}Caswell et al.  1989) to a new
class of galactic head-tail object (\markcite{Bec}Becker \& Helfand 1985; \markcite{Hel}Helfand
\& Becker 1985; see also \markcite{Gra}Gray 1994).  Another intriguing suggestion is that
G357.7$-$0.1 is powered by an object ejected from the nearby SNR (G359.0-0.9) which lies only
$1\arcmin$ from the symmetry axis of G357.7$-$0.1.  This scenario, however, would require such
a `runaway' pulsar or X-ray binary to have a transverse velocity $\ga 2,000$ \kms\/ which is
considered unlikely (\markcite{Gra}Gray 1994).  The discovery of a OH (1720 MHz) maser
coincident with G357.7$-$0.1 has renewed speculation that it is, in fact, a galactic SNR
(\markcite{Fra96}Frail et al.  1996).

The odd morphology of G357.7$-$0.1 has earned it the name `Tornado'.  The reason for this
moniker can be seen in the greyscale continuum images of G357.7$-$0.1 shown in Figure 11 (see
also, \markcite{Shaa}Shaver et al.  1985a; \markcite{Bec}Becker \& Helfand 1985).  Figure 11a
was created from archival VLA data at $\sim 6$ cm from both C and D configuration data.  The
resulting image has a resolution of $13.4\arcsec \times 6.7\arcsec$ (P.A.  = 23.8\arcdeg) and
an rms of $\sim 0.6$ mJy beam$^{-1}$.  The data used to create the image in Figure 11b were
compiled from 18 cm and 20 cm VLA archive data from all four VLA configurations.  The resulting
image is one of the most sensitive and highest resolution images of this source to date with an
rms of $\sim 1.5$ mJy beam$^{-1}$ and resolution of $11\arcsec.6 \times 8\arcsec.1$ (P.A.  =
26.2\arcdeg; see also \markcite{Shaa}Shaver et al.  1985a; \markcite{Bec}Becker \& Helfand
1985; \markcite{Yus99b}Yusef-Zadeh 1999b).  Comparison of the peak flux densities in Figs.  11a
($S_{6{\rm cm}}=$77 mJy beam$^{-1}$) and 11b ($S_{20{\rm cm}}=$145 mJy beam$^{-1}$) confirms
that the spectral index of G357.7$-$0.1 is $\approx -0.5$.

%\clearpage

\begin{figure}[ht]
\epsfxsize=8.5cm \epsfbox{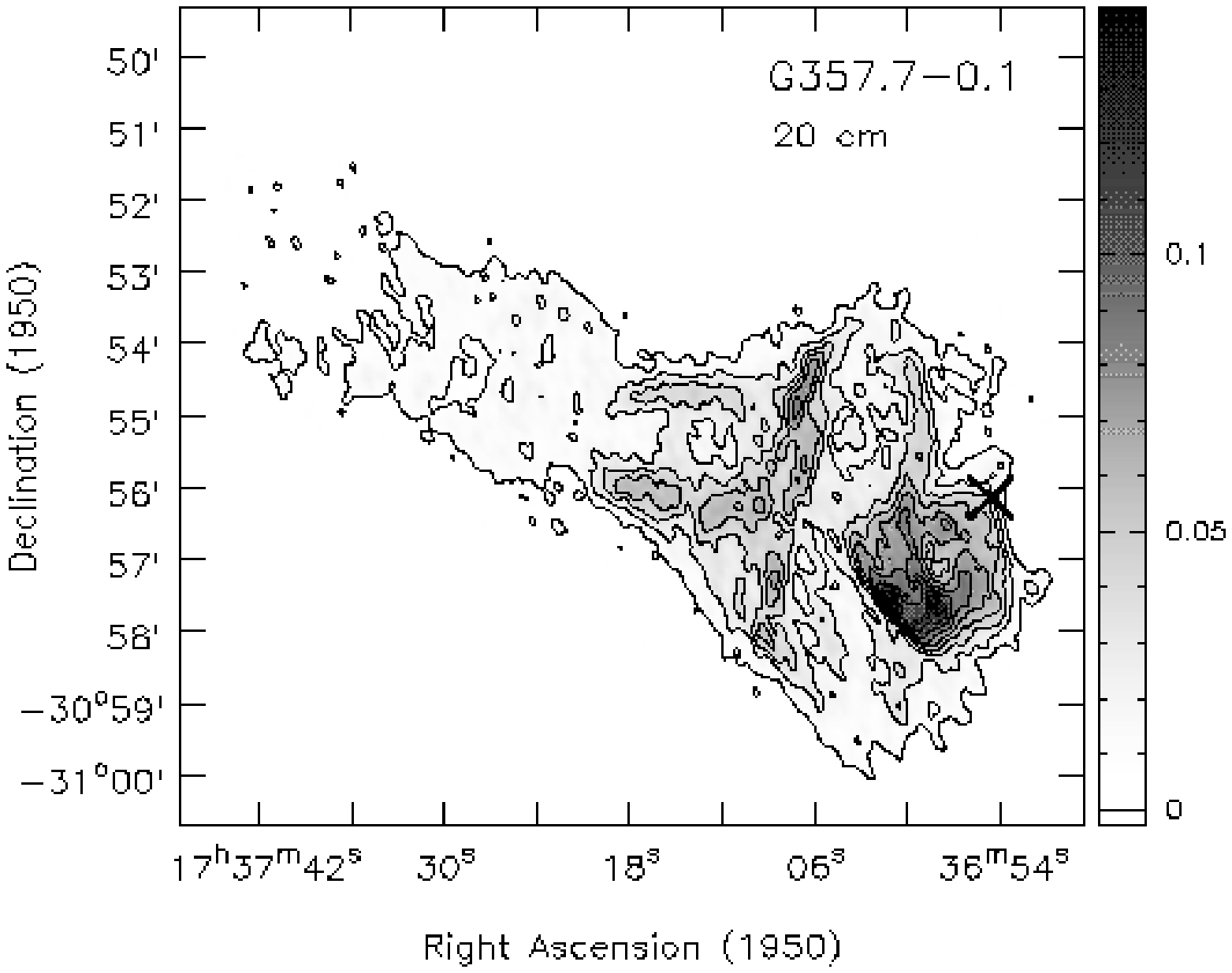}
\figcaption[Filename]{\footnotesize Similar to Figure 11b, but with contours at (0.05, 0.15, 0.25, 
0.35, 0.45, 0.55, 0.65, 0.75, 0.85) $\times$ 145 mJy beam$^{-1}$. The location of 
G357.7-0.1 OH (1720 MHz) maser feature (1) is marked by an ($\times$) symbol\label{figu12}.}
\end{figure}

\begin{figure}[ht]

\epsfxsize=5.0cm \epsfbox{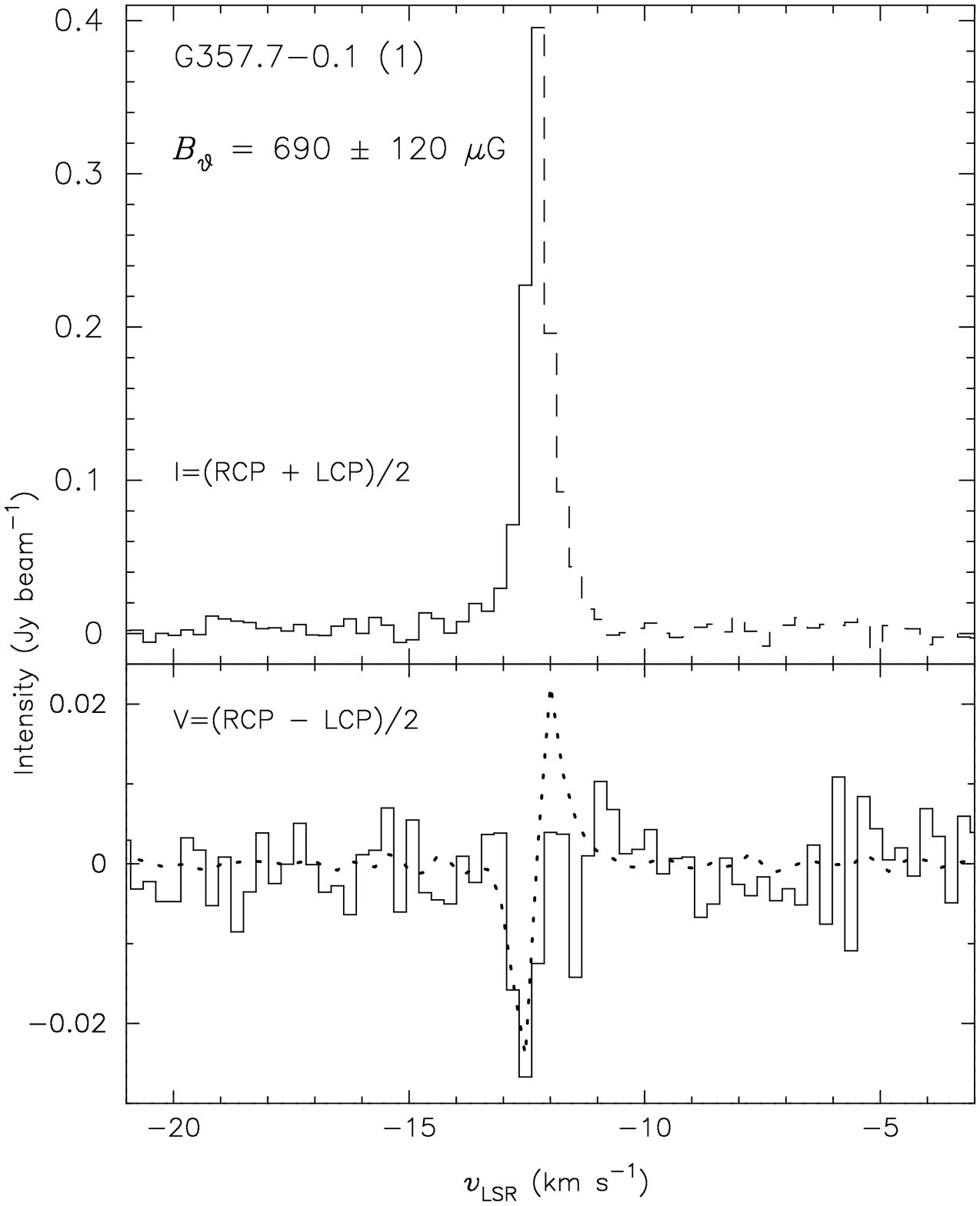}
\figcaption[Filename]{\footnotesize Fit of \Bth\/ for G357.7$-$0.1 OH (1720 MHz) maser feature (1).  
The upper
panels show the VLA Stokes I profiles ({\em solid histogram}), and the bottom panels show the VLA
Stokes V profiles ({\em solid histogram}) with the fitted derivative of Stokes I shown as smooth
dotted curves.  The value of \Bth\/ fit for each position and its calculated error are given at the
top of each plot.  The solid portion of the Stokes I histogram ({\em upper panel}) shows the
velocity range used in the fit\label{figu13}.}
\end{figure}
\clearpage

A compact source at the western edge of G357.7$-$0.1 is also apparent in Figs.  11a and b (also
observed by \markcite{Shab}Shaver et al.  (1985b); see also \markcite{Ste}Stewart et al.  1994,
and references therein).  We obtain flux densities of $70\pm 4$ mJy and $69\pm 12$ mJy for this
compact source at 6 cm and 20 cm respectively, confirming that this source has a flat spectrum
as suggested by \markcite{Shab}Shaver et al.  1985b.  At the resolution of the 18 cm A-array
data ($4\arcsec.5 \times 3\arcsec.4$) this source is extended and has a diameter of $\sim
6\arcsec$ with a possible extension to the east (toward the SNR).  The exact nature of this
source remains unknown.  \markcite{Shab}Shaver et al.  (1985b) cite its flat spectral index at
radio wavelengths as evidence that it is an unrelated \HII\/ region.  Alternatively,
\markcite{Shu}Shull, Fesen, \& Saken (1989) suggest that it could be a pulsar which is
interacting with the SNR shell and is responsible for the peculiar morphology of G357.7$-$0.1
similar to CTB 80.  Higher resolution continuum observations or recombination line observations
of this compact source are needed to determine its nature.

\placefigure{figu12}
\placefigure{figu13}

Figure 12 shows the 20 cm image (similar to Fig.  11b) with the single G357.7$-$0.1 OH (1720 MHz)
maser superposed.  The fit of this maser's \Bth\/ $= 0.7$ mG is shown in Figure 13.  For this maser
feature, only the high velocity side of the line could be adequately fit by the derivative of Stokes
I.  As mentioned for CTB33 OH(1), a suppression of part of the Stokes V signal can be due to the
presence of more than one velocity component, but the narrowness of this line (\dv\/$\sim 0.6$
\kms\/) makes either an inability to fully resolve the Stokes V pattern or poor S/N more probable
explanations.  In the absence of better data, this detection should be regarded as tentative.
\markcite{Yus99b}Yusef-Zadeh et al.  (1999b) also observed G357.7$-$0.1 at 1720 MHz with the VLA in
A-configuration during an OH (1720 MHz) survey of the Galactic center.  With their short integration
time (rms $\sim 12$ mJy), they were able to place an upper limit on \Bth\/ for G357.7$-$0.1 OH(1) of
2 mG, consistent with our 0.7 mG detection.  In addition, these authors detected extended OH (1720
MHz) emission toward G357.7$-$0.1 with the VLA in D-configuration ($114\arcsec \times 38\arcsec$
resolution).  These authors note that this extended emission may originate from low-gain masers
given the lack of OH (1720 MHz) absorption toward this source.

Linear polarization images of the synchrotron emission from G357.7$-$0.1 at 5.8 GHz by
\markcite{Ste}Stewart et al.  (1994) show that the plane-of-sky magnetic field vectors lie
circumferentially to the vertical bands of continuum emission (best seen in Fig.  11).  These
authors note that this morphology is suggestive of a spiral magnetic field structure in the
SNR.  It is also interesting to note that the \markcite{Ste}Stewart et al.  images indicate a
lack of linear polarization at the location of the OH (1720 MHz) maser, as might be expected if
$\vec{B}$ were nearly along the line of sight.

\section{DISCUSSION}

\subsection{Implications from 1720 MHz Maser Theory}

We have assumed until now that magnetic fields can be calculated from the Zeeman effect in masers using
the same formalism that has been developed for thermal radiation (i.e.  \markcite{Tro}Troland \& Heiles
1982 and \Bth\/ in Table 3).  However, stimulated emission is intrinsically different from spontaneous
emission or absorption (see \markcite{Eli96}Elitzur 1996), so it is not unreasonable to imagine that a
maser's polarization properties are different from those of thermal emission.  A common variable in any
polarization study is the ratio of the line splitting ($\Delta\nu_B=Z\mid\vec{B}\mid$) to the Doppler
width ($\Delta\nu_D$) which we will denote $x_B$.  In what follows we assume this ratio is small ($x_B
<< 1$) since we do not resolve the line splitting in our Stokes I spectra (i.e.  the splitting is only
detected in Stokes V).  In this section we review the results of several theoretical maser polarization
studies for the case $x_B<<1$ (\markcite{Eli96}\markcite{Eli98}Elitzur 1996, 1998; and
\markcite{Ned92}Nedoluha \& Watson 1992) and compare the implications and predictions of these studies
to the currently available OH (1720 MHz) Zeeman data.  These comparisons have not proven sufficient to
discern exactly how the total maser magnetic field strength \Bvm\/ scales with our thermal estimates
\Bth\/ (Table 3), but have led to a number of related conclusions which are described below.

One recent analytical study of the polarization properties of masers by \markcite{Eli98}Elitzur (1998)
has suggested that the usual thermal Zeeman formulation, V $=ZC$\Bvm\/\thinspace{$d$I$/d\nu$} with $C$ a
constant, may not be valid for unsaturated masers.  Unfortunately, a key uncertainty in the study of
masers is that their degree of saturation cannot be observed directly.  However, \markcite{Eli98}Elitzur
(1998) also proposes that a maser's saturation state can be revealed by the ratio of its Stokes V
profile to the derivative of Stokes I (i.e.  \Ra\/$=$V/($d$I$/d\nu$)).  This is a consequence of the
fact that maser amplification during unsaturated maser growth is exponential, causing a narrowing of the
maser line.  Under these circumstances, Elitzur (1998) predicts that \Ra\/$=ZC$\Bvm\/ is not a constant
(i.e.  $C \neq$ constant) and instead assumes a Gaussian absorption shape.

A test of this prediction was performed for the \Ra\/ profiles of W51C feature 2 (dynamic range = 720:1)
and the highest dynamic range maser in the \markcite{Cla97}Claussen et al.  (1999) VLBA study of W28
(W28 F39 [A]; dynamic range = 60:1).  Note, that the VLBA OH maser data have three times the spectral
resolution of the VLA data.  The \Ra\/ profiles for both of these masers are relatively flat (to within
their rms) and show no evidence of the Gaussian absorption shape predicted for unsaturated maser
emission.  We conclude however, that while these \Ra\/ profiles are indicative of saturated maser
emission, this test is somewhat impractical for real data.  The derivative of I passes through zero at
line center by definition, and toward the line wings as well depending on the rms of the spectra.  A
more useful variation of this test is to investigate the linear dependence of Stokes V as a function of
$d$I$/d\nu$ since this test does not suffer from the discontinuities that are inevitable with the \Ra\/
test.  The resulting V vs.  $d$I$/d\nu$ plots reveal the straight lines expected from saturated emission
(i.e.  $ZC$\Bvm\/= constant) for W51C OH(2) and W28 F39 [A] along with several more of the masers in
this study (i.e.  G349.7$+$0.2 OH(3), CTB37A OH(1-5)), but remain inconclusive for the lowest S/N masers
(i.e.  CTB33 OH(1)).

Another perspective on a maser's degree of saturation can be gained by comparing an upper limit for its
optical depth ($\tau_{max}$) assuming that it {\em is} unsaturated, with a lower limit ($\tau_{min}$)
based on the flatness of its observed \Ra\/ ratio.  For unsaturated masers, $T_b = T_xexp(\tau_{max})$
(where $\tau_{max}$ is the maser's optical depth at line center and $T_x$ is the excitation temperature;
see \markcite{Eli98}Elitzur 1998 for details).  As mentioned above, the lower limit depends on the
confidence with which a flat \Ra\/ profile can be determined, so that $\tau_{min} =
\epsilon^{-1}$ln\Da\/, where $\epsilon$ is the dynamic range of the observations and \Da\/ is
$\sigma_{B_{\theta}}/$\Bth\/ (\markcite{Eli98}Elitzur 1998).  \markcite{Eli98}Elitzur (1998) notes that
whenever these two limits are inconsistent ($\tau_{min} > \tau_{max}$), the maser must be saturated.
This `inconsistency' was found to be true for every maser in our sample, indicating that OH (1720 MHz)
masers may indeed be saturated.  This was also found to be the case in a OH (1720 MHz) maser study of
the galactic center by \markcite{Yus99a}Yusef-Zadeh et al.  (1999a).  We point out, however, that we
have only been able to calculate a lower limit for $T_{b}$ since the maser spots are not resolved at the
present resolutions.  This means that if the true values of $T_{b}$ (and hence $\tau_{max}$) are large
enough, the apparent saturated maser inconsistencies ($\tau_{min} > \tau_{max}$) could disappear.
However, the sizes found for OH (1720 MHz) masers in the W28 and W44 VLBA observations of Claussen et
al.  (1999) $\sim 50 - 180$ mas, if typical, imply that the majority of masers in this study would still
have $\tau_{min} > \tau_{max}$ as required for saturated maser emission.

Even if these masers are likely to be saturated, a second intrinsic uncertainty in performing Zeeman
analysis on OH masers is the value that should be taken for the constant ``$C$'' in V$ =
ZC$\Bvm\/\thinspace{{\rm I}$'$}.  According to \markcite{Eli98}Elitzur (1998), this constant is modified
from its thermal value:  $C_{th}=cos\thinspace{\theta}$ to $C_m=8/(3pcos\thinspace{\theta})$ for
saturated masers, where $\theta$ is the angle between the magnetic field vector and the line-of-sight.
This solution to the maser polarization problem takes into account the fact that a photon generated via
stimulated emission does not necessarily have the same polarization as the parent photon, and that the
radiation is beamed.  The parameter $p$ in this formulation depends on the geometry of the masing
region, with $p=1$ or 2 for filamentary or planar geometry respectively.  Notice that in addition to the
factor of $8/(3p)$ difference between the thermal and maser equations for $\mid\vec{B}\mid$, they also
have a completely different dependence on $cos{\thinspace \theta}$.  That is, while thermal radiation
samples \Blos\/$=\mid\vec{B}\mid cos{\thinspace \theta}$, masers sample
\Bm\/$\propto\mid\vec{B}\mid/cos{\thinspace \theta}$ according to the Elitzur model.  Therefore, this
model suggests that the magnetic field needed to produce a maser's Stokes V signal is {\em less} than
the field needed to produce a comparable signal from thermal radiation (i.e.  $C_m > C_{th}$).  For
example, using \Bvm\/ = (\Bth\//$C_m$), the Elitzur (1998) model predicts \Bvm\/ = $0.3p$\Bth\/ for
$\theta = 35\arcdeg$, and \Bvm\/ = $0.1p$\Bth\/ for $\theta = 75\arcdeg$.

Ideally, the Stokes parameters U, Q, and V should only depend on $\theta$ and \Bvm\/.  Therefore, if
analytic expressions for the Stokes parameters are known, it should be possible to calculate
$\theta$ and \Bvm\/ explicitly.  According to \markcite{Eli96}Elitzur (1996), the analytic
expression for the linear polarization fraction ($\mid q\mid =({\rm Q}^2 + {\rm U}^2)^{1/2}$/I) has
the approximate form 
\begin{equation}
q\approx \left[1- \left(\frac{2}{3\thinspace{sin^2\thinspace{\theta}}}\right)\right], 
\end{equation}
as long as $x_B << 1$.  Therefore, if the linear polarization fraction is known, Eq.  [1] can
provide an estimate for $\theta$ and hence the magnitude of $C_m$.  For the case of W51C OH(2),
where we have measured $q=3.5$\% (\S 3.2.1), we obtain $\theta \approx 56\arcdeg$ and $C_m\approx
5/p$.  Consequently, in the Elitzur model, W51C OH(2) has $\mid\vec{B}\mid\approx (0.4 \times p)$ mG
compared to its estimated thermal \Bth\/ value of 1.9 mG (where
$p =$ 1 or 2).

Unfortunately, this technique can only produce accurate results if the ``total'' linear polarized
intensity is measured.  The more extensive Stokes Q and U observations of OH (1720 MHz) masers in W28
and W44 (\markcite{Cla97}Claussen et al.  1997), indicate that $q$ only ranges from 0.5\% to 22\% for
all 30 of the masers with positive Q and U detections.  In fact, the linear polarization of the masers
in W28 and W44 is quite low with average values of $4\pm 5$\% and $10\pm 5$\% respectively , similar to
that found for W51C OH(2) ($\pm$ gives the standard deviation of the distribution).  It would be
quite remarkable if $q$, and hence the angle between the line-of-sight and the magnetic field
($\theta$), were so similar in three different SNRs.  Indeed it would be more natural for the polarized
intensity of these masers to be more widely varied as is the case for SiO masers (see for example
\markcite{MacI}McIntosh \& Predmore 1993).  There are at least two other effects in addition to the
viewing angle ($\theta$) which can decrease the magnitude of $q$.  These are 1) Faraday depolarization
within the masing region and 2) curvature of the magnetic field lines within the masing region causing
cancellation within the beam (i.e.  MHD waves, see \markcite{Eli92}Elitzur 1992).

Faraday rotation can only be a significant factor if the Faraday rotation sizescale is smaller
than the length scale of the masing region (see \markcite{Eli92}Elitzur 1992 for details).
From \markcite{Eli92}Elitzur (1992) the Faraday rotation sizescale is $\ell_F=2\times
10^{16}(\lambda^2n_e$\Blos\/)$^{-1}$ where $\lambda$ is the observing frequency in cm, $n_e$ is
the electron density in \cc\/, and \Blos\/ is the magnitude of the magnetic field along the
line of sight in Gauss.  Theoretical modeling of OH (1720 MHz) masers by \markcite{Loc}Lockett
et al.  (1999) and Wardle (1999) suggest that the ionization fraction needed to produced a
large enough column of OH for strong maser action is in the range $10^{-7}\lesssim
\chi_e\lesssim 10^{-5}$.  These authors also find that the density in the masing region must
lie in the range $1\times 10^4$ \cc\/ $\lesssim n_{H_2} \lesssim 5\times 10^5$ \cc\/.  Given
our average magnetic field detection of \Bth\/$\sim 1$ mG, the Faraday rotation length scale
can be written
\footnotesize\begin{equation}\footnotesize
\ell_F=6.2\times 10^{17}\left[ \left(\frac {n_{H_2}}{10^5~{\rm cm^{-3}}}\right)\left(\frac 
{\chi_e}{10^{-6}}\right)\left(\frac {B_{los}}{1~{\rm mG}}\right)\right] ^{-1}~{\rm cm}.
\end{equation}\small
This $\ell_F$ estimate is only three times larger than the OH (1720 MHz) maser gain length
estimated by \markcite{Loc}Lockett et al.  (1999) of $\sim 2\times 10^{17}$ cm, implying that Faraday
depolarization may contribute to the low linear polarization intensities observed by
\markcite{Cla97}Claussen et al (1997).  The decrease of polarization intensity with increasing $J$ in
SiO masers is also thought to be a consequence of such Faraday depolarization (\markcite{MacI}McIntosh
\& Predmore 1993; or \markcite{Wall97}Wallin \& Watson 1997 for an opposing view).  Unfortunately,
the magnitude of such depolarization is difficult to quantify and quite model dependent (see e.g.
Wallin \& Watson 1997).

In addition, some evidence of tangling in the magnetic field lines may be indicated from the
ten-fold increase in \Bth\/ measured toward W28 F39 [A] with the VLBA ($\sim 2$ mG) compared to
the average value measured throughout W28 with the VLA ($\sim 0.2$ mG; \markcite{Cla99}Claussen
et al.  1999).  However, this particular spot did not have a positive VLA \Bth\/ detection (due
to the presence of multiple spatial components at the lower resolution) so a direct comparison
is not possible.  More VLA vs.  VLBA magnetic field strength comparisons are needed to
determine whether this apparent increase in \Bth\/ with higher resolution is real and/or
common.

These examples suggest that the linear polarization intensities observed toward W51C, W28, and W44 could
be reduced from their `true' values by one or both of these effects, making the calculation of $\theta$
directly from $q$ unreliable.  Alternatively, using the condition that $q^2 + v^2 \leq 1$, (where
$v=$V/I), it is possible to obtain the following constraint on $\theta$ at the frequency of the peak in
Stokes V, $16(x_B)^2\lesssim cos^2\thinspace{\theta}\lesssim 2/3$ as long as $x_B < 0.2$ (see Elitzur
1996 for details).  This means that the angle between $\vec{B}$ and the line of sight must be greater
than $\sim 35\arcdeg$ in order for polarized emission to be observed at all (see also
\markcite{Gol}Goldreich, Keeley, \& Kwan 1973).  This lower limit on $\theta$ ($\sim 35\arcdeg$),
suggests that the magnetic field values reported in Table 3 (\Bth\/) are {\em overestimated} by factors
of at least 1.5 ($p=2$).  The upper bound on $\theta$ cannot be accurately calculated from these data
since it depends on $x_B$ ($Z$\Bth\//$\Delta\nu_D$) and our current spectral resolution (0.27 \kms\/;
the highest available with the VLA in 2IF mode) is insufficient to resolve line splittings with $x_B <
0.2$.  However, since \Bth\/ is an {\em upper limit} on the total field strength (\Bvm\/), a useful
limit on $x_B$ can be made by taking the average value of \Bth\//\dv\/ using data from Tables 2 and 3,
in which case $x_B\lesssim 0.12$ on the average and $\theta\lesssim 61\arcdeg$.  This upper limit to
$\theta$ suggests that the average total field strength is \Bvm\/$\gtrsim 0.2p$\Bth\/, i.e.  the thermal
approximation (\Bth\/ in Table 3) {\em overestimates} the maser field (\Bvm\/) by less than a factor of
$C_m\sim 5/p$.

It is important to note, however, that the analytic expressions formulated by Elitzur are controversial
and that other maser studies have reached different conclusions.  For example, the numerical simulations
of \markcite{Ned92}Nedoluha \& Watson (1992) suggest that if $x_B << 1$ and the maser is not {\em
strongly} saturated, the thermal Zeeman relationship is a good approximation (i.e.  $C_m\sim 1$).
Unfortunately, it is quite difficult to compare the two methods since the Elitzur model is an analytical
treatment, while the Nedoluha \& Watson (1992) result arises from a numerical simulation.  Also notable
is the \markcite{Ned90}Nedoluha \& Watson (1990) finding that it is very difficult to produce linear
polarization in masers unless they are at least {\em partially saturated}.  It is impossible to make any
quantitative statements on this issue with the data presented here since we only attempted linear
polarization measurements toward one of our sources (W51C).  However, the \markcite{Cla97}Claussen et
al.  (1997) OH (1720 MHz) maser study of all four Stokes parameters toward the SNRs W28, W44, and IC443
contains a total of 49 masers with $S_{\nu}>200$ mJy.  About $60\%$ of these have positive Q and/or U
detections.  In addition, all 13 of the Claussen et al.  (1997) masers with $3\sigma$ \Bth\/ detections
also have positive Stokes Q and/or U detections.  These comparisons suggest that in the Nedoluha \&
Watson model strong OH (1720 MHz) masers are also likely to be at least partially saturated.

An additional observational test offered by maser theory is to check whether there is any correlation
between the linear polarization position angle (P.A.)  and the SNR shock front in the vicinity of a
maser spot.  Several studies have suggested that OH (1720 MHz) masers must arise in shocks which are
propagating transverse to the line of sight (\markcite{Loc}Lockett et al.  1999; \markcite{Wardle}Wardle
1999; \markcite{Fra98}Frail \& Mitchell 1998; \markcite{Cla97}Claussen et al.  1997).  It is also likely
that the magnetic field vector lies preferentially in the plane parallel to the shock front since only
this component can be amplified by shock compression.  In addition, the linear polarization P.A.  for
masers, unlike that of thermal dust emission, can be either parallel or perpendicular to the direction
of the magnetic field in the plane of the sky.  This difference occurs because the asymmetry of dust
grains removes one degree of freedom.  For these reasons, we might expect the observed P.A.  of the
linear polarization to be either parallel or perpendicular to the shock front.  Of course such a
comparison is only meaningful if significant Faraday rotation of the P.A.  has not occurred either
because the Faraday depolarization is insignificant or because the P.A.  of the linear polarization is
unaffected by it.  A lack of {\em net} P.A.  rotation in the presence of depolarization is suggested by
Elitzur (1992) and the SiO maser observations of McIntosh \& Predmore (1993), although Wallin \& Watson
(1997) present an alternate viewpoint based on numerical simulations.

For the case of W44 region E, the linear polarization angle is $\sim -28\arcdeg$
(\markcite{Cla97}Claussen et al.  1997) and the OH (1720 MHz) masers are also distributed along a line
which is oriented NW/SE at about the same angle.  Further indication of the orientation of the shock
front comes from CO ($J=3-2$) observations by \markcite{Fra98}Frail \& Mitchell (1998) which show a
ridge of shocked CO emission parallel to the line of masers.  This example suggests that there is a
correlation between OH (1720 MHz) linear polarization position angles and the orientation of the shock
front.  The presence of such a correlation could argue against the possibility of significant Faraday
rotation within the SNRs or may result from Faraday depolarization without significant net rotation.
In any case, additional linear polarization measurements in conjunction with molecular observations of OH
(1720 MHz) maser regions are needed to confirm this connection.

These comparisons between observation and theory of OH (1720 MHz) masers have led to the following
conclusions:  (1) It is likely that OH (1720 MHz) masers are saturated based on their flat \Ra\/
profiles (determined directly and from estimates of maser optical depths; \markcite{Eli98}Elitzur 1998).
This conclusion is also supported by the high incidence of linear polarization in these masers
(\markcite{Ned92}Nedoluha \& Watson 1992; \markcite{Cla97}Claussen et al.  1997).  (2) The linear
polarization intensities of OH (1720 MHz) masers (needed to calculate \Bvm\/ in the Elitzur model) may
be significantly reduced from their intrinsic values by Faraday rotation and/or by tangling in the
magnetic field lines.  This also means that it is impossible to calculate the reduction of \Bvm\/
compared to \Bth\/ using these data directly (\Bvm\/ $\propto$ \Bth\/$cos{\thinspace \theta}$).
However, for angles between 35$\arcdeg$ and 75$\arcdeg$, the Elitzur model predicts that the total
magnetic field strength should range from \Bvm\/ = (0.3 $-$ 0.1)p\Bth\/ (where $p=1$ or 2 for
filamentary or planar geometry respectively).  Alternatively the numerical simulations of Nedoluha \&
Watson (1992) suggest that this correction factor is essentially unity.  (3) It may be possible to predict the
direction of the shock in which an OH (1720 MHz) maser arises from the position angle of its observed
linear polarization (as long as the P.A. is not significantly rotated by Faraday rotation).

\subsection{Magnetic Fields in Shocked Molecular Gas} 

In this study we have reported ten new measurements of OH (1720 MHz) Zeeman magnetic field strengths in
five galactic SNRs.  Previous studies of this type (\markcite{Cla97}Claussen et al.  1997;
\markcite{Kor}Koralesky et al.  1998; \markcite{Yus96}Yusef-Zadeh et al.  1996) have measured \Bth\/ in
an additional five SNRs.  The magnitude of \Bth\/ in all of these SNRs (including those measured here)
have typical values between 0.1 and 4 mG.  However, as the discussion in \S 4.1 demonstrates, the
conversion factor between thermal estimates of \Bth\/ and the true maser field \Bvm\/ is uncertain.
Therefore, although we will continue to reference our measured fields \Bth\/ (Table 3), it should be
kept in mind that \Bth\/ could be an upper limit that overestimates the true field by less than a factor
of five.

Maser theory suggests that OH (1720 MHz) masers originate in shocked molecular clouds (e.g.
\markcite{Loc}Lockett et al.  1999; \markcite{Wardle}Wardle 1999).  If so, the measurements
reported here of \Bth\/ (and in the references cited above) must originate in post-shock gas.
These theories suggest that OH (1720 MHz) masers can only be pumped efficiently for densities
in the range $1\times 10^4$ \cc\/ $\lesssim n_{H_2} \lesssim 5\times 10^5$ \cc\/ and
temperatures in the range 50 K $\lesssim T \lesssim$ 125 K (see \markcite{Loc}Lockett et al.
1999).  Indeed, when independent measurements of the conditions in the post-shock gas have been
made (see \markcite{Fra98}Frail \& Mitchell; \markcite{Rea98}\markcite{Rea99}Reach \& Rho 1998,
1999) the gas properties are in agreement with these theoretical expectations.

It remains an open question how magnetic fields of these strengths are generated; i.e.  shock
compression vs.  turbulent amplification (\markcite{Jun}Jun \& Norman 1996).  In what follows
we will show that compression of the existing ambient molecular cloud field is all that is
required to produce the observed field strengths.  One further argument against significant
turbulent amplification of the fields is the likelihood of destroying the maser action due to
loss of velocity coherence in a turbulent velocity field.

A number of Zeeman studies have been undertaken in the past decade to detect magnetic fields in
molecular clouds for the purpose of studying star formation (e.g.  \markcite{Rob}Roberts et al.
1993; \markcite{Bro} Brogan et al.  1999).  In a recent review by \markcite{Cru}Crutcher (1999)
of the existing data for star forming regions, the magnetic field was found to scale with
density as $\mid\vec{B}\mid \propto n^{0.47}$.  \markcite{Cru}Crutcher notes that there are two
possible physical interpretations for this relationship:  (1) Such a relationship between
$\mid\vec{B}\mid$ and $n$ has been predicted by \markcite{Fie}Fiedler \& Mouschovias (1993)
based on studies of ambipolar diffusion; (2) A similar relation is suggested by the observed
invariance of the Alfv\'enic Mach number $m_A = \sqrt 3\sigma/V_A \approx 1$ in molecular
clouds, where $V_A=\mid\vec{B}\mid/4\pi\rho^{1/2}$ and $\sigma = \Delta v/(8\ln 2)^{1/2}$ (see
\markcite{Ber}Bertoldi \& McKee 1992; \markcite{Zwe}Zweibel \& McKee 1995).  This invariance
implies $\mid\vec{B}\mid \propto \Delta v\sqrt\rho$.

\begin{figure}[ht]
\epsfxsize=8.5cm \epsfbox{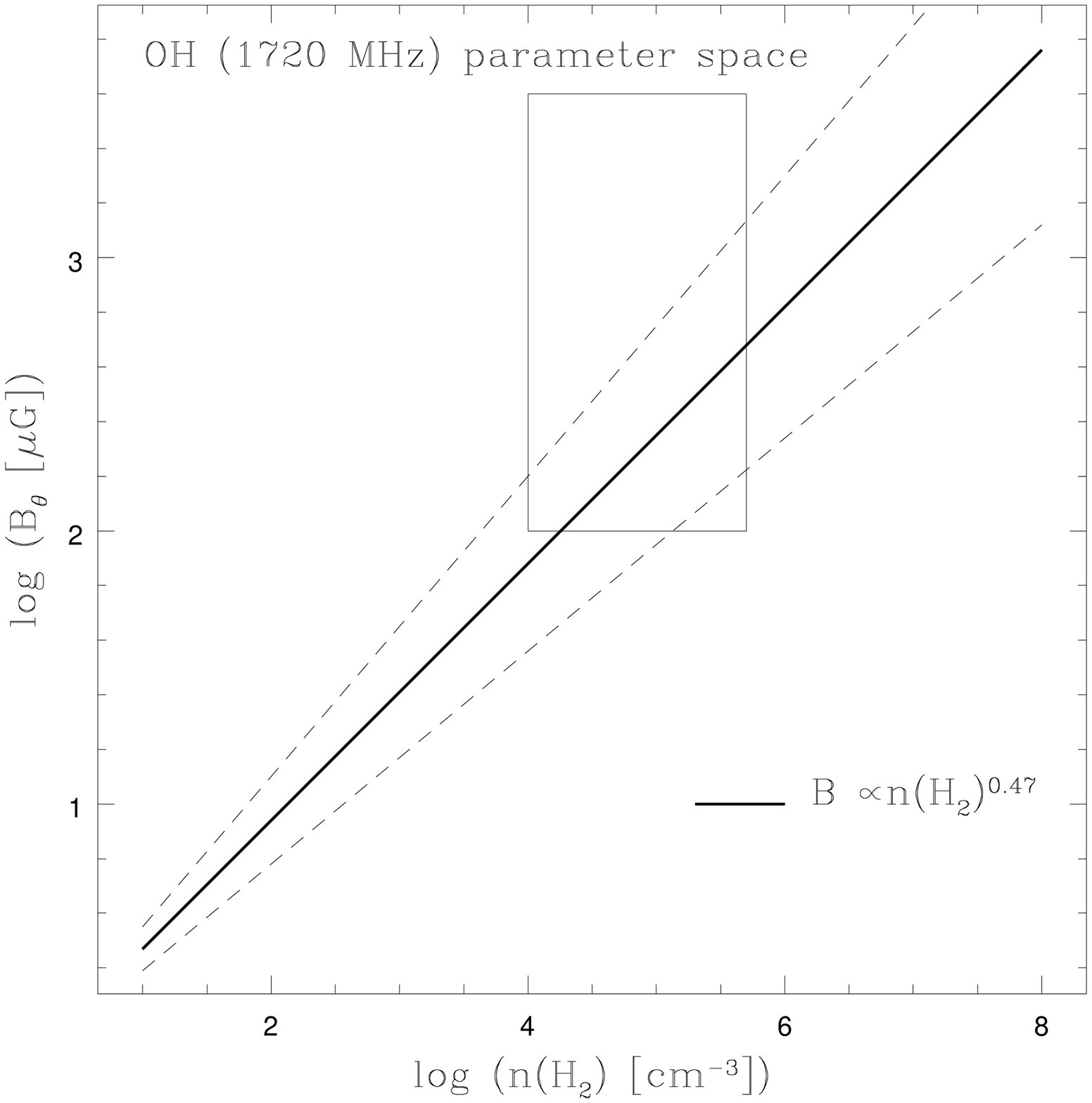}
\figcaption[Filename]{\footnotesize Plot of \Bth\/$ \propto n^{0.47}$ (Crutcher 1999; solid thick line) with the 
OH (1720 MHz) maser's \Bth\/ parameter space superposed.  The two dashed lines show the $1\sigma$ 
errors on the fit obtained by Crutcher\label{figu14}.}
\end{figure}

\placefigure{figu14}

For the range of gas densities expected in OH (1720 MHz) maser regions ($1\times 10^4$ \cc\/
$\lesssim n_{H_2} \lesssim 5\times 10^5$ \cc\/; \markcite{Loc}Lockett et al.  1999), the range
of magnetic fields predicted by the relation $\mid\vec{B}\mid \propto n^{0.47}$ is 75 \muG\/ to
475 \muG\/.  Clearly the OH (1720 MHz) \Bth\/ measurements reported in this work (Table 3) are
greater than those predicted by \markcite{Cru}Crutcher's relation (see also Figure 14).
However, there is a great deal of uncertainty in this statement since \markcite{Cru}Crutcher
considers \Blos\/ (\Blos\/$=\mid\vec{B}\mid cos{\thinspace \theta}$) which {\em underestimates}
the field by factors of $\sim 2$ while our determination of \Bth\/ could {\em overestimate} the
field by as much as a factor of five according to the Elitzur model (see \S 4.1).

If we assume, however, that the two types of magnetic field measurements (molecular cloud vs.
OH (1720 MHz) masers) are approximately comparable (i.e.  the thermal Zeeman equation is valid
for masers) it may not be surprising that their magnitudes are different.  This is because
theory and observations suggest that OH (1720 MHz) masers do not originate in undisturbed
``normal'' molecular clouds, but rather have experienced a shock (see above).  If this is the
case, the \markcite{Ber}Bertoldi \& McKee interpretation for the scaling of $\mid\vec{B}\mid$
($\propto \Delta v\sqrt\rho$) suggests that OH (1720 MHz) masing regions simply have larger
linewidths for a given density than unshocked molecular clouds.  If we set $V_A=\sqrt 3\sigma$
as suggested by \markcite{Ber}Bertoldi and McKee (1992), then $\mid\vec{B}\mid=0.4 \Delta v
n_{p}^{1/2}$ where $n_{p}$ is the proton density in \cc\/ and the line width ($\Delta v$) is in
\kms\/.  For the case of the W51C OH (1720 MHz) masers, we can take $\Delta v = 10$ \kms\/
based on the CO observations of \markcite{Koo99}Koo (1999), and $n_{p} = 1\times 10^5$ \cc\/
from the typical density needed to excite OH (1720 MHz) masers (\markcite{Loc}Lockett et al.
1999; Wardle et al.  1999).  With these estimates, the implied magnetic field strength is
\Bvm\/$=1.3$ mG, in close agreement with the \Bth\/ values observed toward W51C.

Alternatively, if the \Bth\/'s measured in OH (1720 MHz) masers should be reduced by factors of
$\lesssim 5$ (i.e.  Elitzur's maser polarization model), the fields estimated from OH (1720 MHz)
masers are in agreement with the values predicted by \markcite{Cru}Crutcher's relation.  In
this case, the magnetic fields in masing regions and molecular clouds follow the same scaling
with density without shock amplification of the field, i.e.  no $\Delta v$ dependence.  Such
agreement might be expected if the observed scaling were the result of ambipolar diffusion in
both types of regions.  Clearly convergence on our understanding of the nature of maser
polarization (and hence \Bvm\/) is needed to distinguish between the two possibilities.

In a radiative shock, compression of the gas follows the relation $\eta = \sqrt 2V_s/V_{Ao}$,
where $V_{Ao}$ is the Alfv\'en velocity in the pre-shock gas (see \markcite{Dra}Draine \& McKee
1993).  Using the relation $B_o\propto n_o^{1/2}$, it can be shown that $V_{Ao}\simeq 2$ \kms\/
(see also \markcite{Hei}Heiles et al.  1993).  Thus for reasonable shock velocities (10 - 50
\kms\/; \markcite{Loc}Lockett et al.  1999; \markcite{Fra98}Frail \& Mitchell 1998), $\eta = 7
- 35$.  This compression is sufficient to enhance $B_o$ to \Bth\/ without invoking turbulent
enhancement of the magnetic field, assuming that $B_{ps}/B_o = \eta$ (see
\markcite{Che}Chevalier 1999; \markcite{Fra98}Frail \& Mitchell 1998).  For example, if we take
a typical shock speed of $V_s \sim 25$ \kms\/ and typical values of $B_o \sim 70$ \muG\/ for
molecular clouds of density $n_o \sim 5 \times 10^3$ \cc\/ (\markcite{Fra98}Frail \& Mitchell
1998; \markcite{Wardle}Wardle 1999; \markcite{Cru}Crutcher 1999), $\eta \approx 18$ and
$B_{ps}\sim 1.2$ mG -- in close agreement with the observed values of \Bth\/.  This
demonstration shows that typical cloud and shock parameters can lead to enhancement of the
pre-shock magnetic fields to the observed values of \Bth\/.

\subsection{Energetics Implied by Observed \Bth\/}

Using the thermal Zeeman equation, we have detected magnetic fields (\Bth\/) between 0.2 and 2 mG
(Table 3) in OH (1720 MHz) masers associated with five galactic SNRs.  However, as described in \S
4.1 these thermal field strengths may {\em overestimate} the ``true'' field strength \Bvm\/ by a
factor between $\sim 2-5$ (i.e Elitzur 1998).  Alternatively, other theoretical studies suggest
that \Bth\/ is a good approximation to \Bvm\/ (i.e.  Nedoluha \& Watson 1992).  For this reason,
along with our inability to calculate the exact correction factor in the Elitzur model, we have
chosen to use \Bth\/ in the calculations below. However, the fact that \Bth\/ could be an upper
limit to the magnetic field strength should be kept in mind.

Regardless of the exact details of the \Bth\/ to \Bvm\/ conversion
or the origin of these fields (argued to be shock compression of the ambient field above), it is
clear that the OH (1720 MHz) fields are strong (compared to those found in dark clouds $\sim 30$
\muG\/, for example).  This is an important finding since the magnetic field plays a key role in
many aspects of the SNR/molecular cloud interaction including :  (1) As noted above, the magnetic
field determines and, in fact, limits the amount of shock compression in the post-shock gas.  (2)
Magnetic support may help stabilize the post-shock cloud, via the magnetic force acting on ions
perpendicular to the magnetic field lines.  (3) In the model of \markcite{Loc}Lockett et al.
(1999), heating by ambipolar diffusion is needed to extend the length of time the post-shock gas
spends at temperatures favorable to OH (1720 MHz) maser inversion.  Ambipolar diffusion is also
thought to be the method by which subcritical clouds dissipate their magnetic energy and form stars
(\markcite{Cio}Ciolek \& Mouschovias 1995).  Given the many roles that the magnetic field can play
in the evolution of a shocked cloud, we estimate the magnetic energy compared to other energy
sources within the region below.

From the values of \Bth\/ listed in Table 3 the magnetic pressure ($P_{B_{ps}}=B_{ps}^2/8\pi$), in
the post-shock gas of the five SNRs studied here, ranges from $10^{-7}$ to $10^{-9}$ erg \cc\/.
These values are large compared to both $P_{ISM}$ ($\sim 5 \times 10^{-13}$ erg \cc\/) and
$P_{thermal}$ of the hot X-ray emitting gas interior to the remnant, $\sim 2 \times 10^{-10}$ erg
\cc\/ (\markcite{Kul}Kulkarni \& Heiles 1988; \markcite{Fra98}Frail \& Mitchell 1998;
\markcite{Cla97}Claussen et al.  1997).  From the X-ray observations of \markcite{Koo95}Koo et al.
(1995) of W51C ($n_e=0.3$ \cc\/; $T_e = 3 \times 10^6$ K), we can estimate directly that
$P_{thermal} = 2n_ekT_e \approx 3 \times 10^{-10}$ erg \cc\/ for this SNR.  The ram pressure of the
shock can also be compared to $P_{B_{ps}}$ since they should be approximately equal if the field
strengths are proportional to the amount of shock compression, i.e.  $B_{ps}^2/8\pi = \rho _oV_s^2$.
For example, using the values of $n_o$ and $V_s$ estimated in \S 4.2 ($n_o=5 \times 10^3$ \cc\/ and
$V_s=25$ \kms\/), $\rho _o V_s^2 = 5 \times 10^{-8}$ erg \cc\/.  The equivalent magnetic pressure
for W51C OH(1) (\Bth\/$=1.5$ mG) is $9 \times 10^{-8}$ erg \cc\/, in close agreement with the
estimated ram pressure.  Therefore, the magnetic pressure in the post-shock gas dominates over the
thermal pressure of the SNR but is approximately equal to its ram pressure.  The fact that the ram
pressure is almost three orders of magnitude larger than the thermal pressure of the shock indicates
that this SNR is in the radiative phase as expected for SNRs with OH (1720 MHz) masers.

\section{SUMMARY AND CONCLUSIONS}

We have observed the OH (1720 MHz) line in five galactic SNRs to measure their magnetic field strengths
using the Zeeman effect.  We detected all 12 of the bright ($S_{\nu} > 200$ mJy) OH (1720 MHz) masers
previously observed by \markcite{Fra96}Frail et al.  (1996) and \markcite{Gre}Green et al.  (1997) and
measured significant magnetic fields (i.e.  $ > 3\sigma$) in ten of them.  The estimated field
strengths, which we denote \Bth\/, range from 0.2 to 2 mG and are in good agreement with those measured
in the five other SNR (0.1 - 4 mG) for which Zeeman OH (1720 MHz) maser studies exist (Koralesky et al.
1997 and references therein).  In these studies, the field strengths were calculated using the thermal
Zeeman equation for Zeeman splitting less $<<$ the line width (i.e.  V
$=Z$\Bth\/\thinspace{$d$I$/d\nu$}).  In this formula, the total field strength (\Bvm\/) is related to
\Bth\/ by \Bvm\/=\Bth\//$C_m$ where $C_m$ is a function of the viewing angle $\theta$
($C_m=cos\thinspace{\theta}$ for thermal emission).  However, there exists some controversy about the
$\theta$ dependence of $C_m$ for saturated masers and whether it is even a constant for unsaturated
masers (see \S 4.1).  For example, in the Elitzur (1998) model $C_m\propto 1/(cos\thinspace{\theta})$
for masers.  {\em From comparison of these data with maser polarization studies we conclude that these
OH (1720 MHz) masers are likely to be saturated and that $C_m\lesssim 5$, i.e.  the \Bth\/ values
reported in Table 3, overestimate the actual field strengths (\Bvm\/) by less than a factor of five.}

We estimate that these magnetic field strengths are consistent with the hypothesis that ambient
molecular cloud B-fields are compressed via the SNR shock to the observed values (\S 4.2).  Indeed,
field strengths of this magnitude exert a considerable influence on the properties of the shocked
cloud.  For example the magnetic pressures estimated from the values of \Bth\/ listed in Table 3
($10^{-7} - 10^{-9}$ erg \cc\/) far exceed the pressure in the ISM or even the thermal pressure of
the hot gas interior to the remnant (\S 4.3).  We also find that this range of magnetic pressures is
in very good agreement with the ram pressure expected from C-type radiative shocks.

In \S 4.2 and \S 4.3 we show that there is excellent agreement between our values of \Bth\/ and
those implied by shock compression and ram pressure using typical values from the literature.
It is somewhat difficult to maintain this agreement if \Bth\/ should be reduced by as much as a
factor of $\lesssim 5$ (i.e.  Elitzur's maser polarization model \S 4.1), although there is
probably sufficient uncertainty in the parameters to allow such fine-tuning.  We also show that
the observed values of \Bth\/ for OH (1720 MHz) masers are greater than those observed in
molecular clouds for the same range of densities (\markcite{Cru}Crutcher 1999).  It is possible
that this difference is due the intrinsic physical nature of the $\mid\vec{B}\mid \propto
n^{1/2}$ relation (\S 4.2) or may be the result of overestimating the maser B-fields (\S 4.1).

In the future, knowledge of the field strength could be used in conjunction with molecular data
to study the physics of these molecular shocks in more detail.  In one such study
\markcite{Fra98}Frail \& Mitchell (1998) mapped the distribution of molecular gas in the
vicinity of several masers spots in W28 and W44 with the JCMT.  These observations revealed
that the OH masers are preferentially located along the edges of thin filaments or clumps of
molecular gas, suggesting compression of the gas by the passing shock.  In addition to this
morphological evidence, the density, temperature and velocity dispersion of the gas at these
locations suggested that the OH (1720 MHz) masers originate in post-shock gas.  Combining the
VLA and JCMT data they were able to show directly that the magnetic pressure dominates over the
thermal pressure in the post-shock gas, balancing against the ram pressure of the gas entering
the shock wave (i.e.  B$_{ps}^2/8\pi=\rho_\circ{\rm V}_s^2$).  Thus OH (1720 MHz) measurements
of \Bth\/ lead to constraints on the physics of these molecular shocks that are difficult to
obtain any other way.  When used in conjunction with molecular observations, it should be
possible to fully specify the properties (i.e.  geometry, density, temperature, velocity) of
C-type shocks.

\acknowledgments

C.  Brogan would like to thank NASA/EPSCoR for fellowship support through the Kentucky Space
Grant Consortium, as well as, summer student support from NRAO.  We would also like to thank M.
Elitzur and G.  Nedoluha for useful discussions on maser theory and M.  Claussen for providing
his VLBA W28 data and useful comments on the manuscript. In addition, we thank the anonymous 
referee for many useful suggestions.


\begin{references}\footnotesize

\reference{Bec} Becker, R. H., \& Helfand, D. J. \ 1985, Nature, 313, 115

\reference{Ber} Bertoldi, F., \& McKee, C. F. \ 1992, \apj, 395, 140

\reference{Bro} Brogan, C. L., Troland, T. H., Roberts, D. A., \& Crutcher, R. M. \ 1999, ApJ, 515, 304

\reference{Cas} Caswell, J. L., Kesteven, M. J., Bedding, T. R., \& Turtle, A. J. \ 1989 
Proc. Astron. Soc. Aust., 8, 184

\reference{Che} Chevalier, R. A. \ 1999, ApJ, 511, 798

\reference{Cio} Ciolek, G. E. \& Mouschovias, T. Ch. \ 1995, \apj, 454, 194

\reference{Cla97} Claussen, M. J., Frail, D. A., Goss, W. M., \& Gaume, R. A. \ 1997, ApJ, 489, 143

\reference{Cla99} Claussen, M. J., Frail, D. A., Goss, W. M., \& Desai, K. \ 1999, ApJ, 522, 349

\reference{Cor} Corbel, S., Chapuis, C., Dame, T. M., \& Durouchoux, P. \ 1999, ApJ, in press

\reference{Cru} Crutcher, R. M. \ 1999, \apj, 520, 706

\reference{Dic} Dickel, J. R., van Breugel, W. J. M., \& Strom, R. G. \ 1991, AJ, 101, 2151

\reference{Dra} Draine, B. T., \& McKee, C. F. \ 1993, ARA\&A, 31, 373

\reference{Eli92} Elitzur, M., \ 1992 Astronomical Masers (Dordrecht:Kluwer), 6.8

\reference{Eli96} Elitzur, M. \ 1996, \apj, 457, 415

\reference{Eli98} Elitzur, M. \ 1998, \apj, 504, 390

\reference{Fie} Fiedler, R. A., \& Mouschovias, T. Ch. \ 1993, \apj, 415, 680

\reference{Fra94} Frail, D. A., Goss, W. M., \& Slysh, V. I. 1994, ApJ, 424, L111

\reference{Fra96} Frail, D. A., Goss, W. M., Reynoso, E. M., Giacani, E. B.,  Green, A. J.,
\& Otrupcek, R. \ 1996, AJ, 111, 1651

\reference{Fra98} Frail, D. A., \& Mitchell, G. F. 1998, ApJ, 508, 690

\reference{Gol} Goldreich, P., Keeley, D. A., \& Kwan, J. Y. \ 1973, ApJ, 179, 111 

\reference{Gra} Gray, A. D. \ 1994, \mnras, 270, 835

\reference{Gre} Green, A. J., Frail, D. A., Goss, W. M., \& Otrupcek, R. \ 1997, AJ, 114, 2058

\reference{Hei} Heiles, C., Goodman, A. A., McKee, C. F., \& Zweibel, E. G.\ 1993,  
Protostars and Planets III, ed. E.H. Levy \& J. I. Lunine (Tucson: Unv. of 
Arizona Press), 279167-181

\reference{Hel} Helfand, d. J. \& Becker, R. H. \ 1985, Nature, 313, 118

\reference{Hur} Hurley, K., Kouveliotou, C., Woods, P., Mazets, E., Golenetskii, S., Frederiks, D. D., 
Cline, T., \& Van Paradijs, J. \ 1999, \apj, 519, L143

\reference{Jon} Jones, T. w., Rudnick, L., Jun, B., Borkowski, K. J., Dubner, G., Frail, D. A., 
Kang, H., Kassim, N. E., \& McCray, R. \ 1998, PASP, 110, 125

\reference{Jun} Jun, B. \& Norman, M. L. \ 1996, ApJ, 472, 245

\reference{Kas} Kassim, N. E., Baum, S. A., \& Weiler, K. W. \ 1991, \apj, 374, 212

\reference{Kle} Klein, R. I., McKee, C. F. \& Colella, P. \ 1994, \apj, 420, 213 

\reference{Koo95} Koo, B. -C., Kim, K. -T., \& Seward, F. D. \ 1995, \apj, 447, 211

\reference{Koo97a} Koo, B. -C., \& Moon, D. -S., \ 1997a, \apj, 475, 194

\reference{Koo97b} Koo, B. -C., \& Moon, D. -S., \ 1997b, \apj, 485, 263

\reference{Koo99} Koo, B. -C. \ 1999, ApJ, 518, 760

\reference{Kor} Koralesky, B., Frail, D. A., Goss, W. M., Claussen, M. J, \& Green, A. 
J. \ 1998, AJ, 116, 1323

\reference{Kul} Kulkarni, S. R., \& Heiles, C.  \ 1988, Galactic \& Extragalactic Radio Astronomy 
(2nd edition) Berlin and New York, Springer-Verlag, 95

\reference{Loc} Lockett, P., Gauthier, E., \& Elitzur, M. \ 1999, ApJ, 511, 235

\reference{Lyn} Lyne, A. G., \& Lorimer, D. R. \ 1994, Nature, 369, 127

\reference{MacI} McIntosh, G. C., \& Predmore, C. R. \ 1993, ApJ, 404, L71

\reference{MacL} MacLow, M.-M., McKee, C. F., Klein, R. I., Stone, J. M., \& Norman, M. 
L., \ 1994, ApJ, 433, 757

%\reference{McK92} McKee, C. F., \& Zweibel, E. G. \ 1992, \apj, 399, 551

%\reference{McK95} McKee, C. F., \& Zweibel, E. G. \ 1995, \apj, 440, 686

\reference{Mie} Miesch, M. S. \& Zweibel, E. G. \ 1994, \apj, 432, 622

\reference{Mil} Milne, D.  K.  \ 1990, in Proc.  140th IAU Symp., Galactic and Intergalactic Magnetic
Fields, Beck, B., Kronberg, P.  P., \& Wielebinski, R., ed., 67

\reference{Ned90} Nedoluha, G. E., \& Watson, W. D. \ 1990, \apj, ApJ, 354, 660

\reference{Ned92} Nedoluha, G. E., \& Watson, W. D. \ 1992, \apj, ApJ, 384, 185

\reference{Rea98} Reach, W. T., \& Rho, J. \ 1998, \apj, 507, L93

\reference{Rea99} Reach, W. T., \& Rho, J. \ 1999, \apj, 511, 836

\reference{Rob} Roberts, D. A., Crutcher, R. M., Troland, T. H., \& Goss, W. M. \ 1993, ApJ, 412, 675

\reference{Sar} Sarma, A. P., Goss, W. M., Green, A. J., \& Frail, D. A. \ 1997, 483, 335
 
\reference{Shaa} Shaver, P. A., Salter, C. J., Patnaik, A. R., van Gorkom, J. H. \& Hunt, G. C. 
\ 1985a, Nature, 313, 113

\reference{Shab} Shaver, P. A., Pottasch, S. R., Patnaik, A. R., van Gorkom, J. H. \& Hunt, G. C. 
\ 1985b, A\&A, 147, L23

\reference{Shu} Shull, M. J., Fesen, R. A., \& Saken, J. M. \ 1989, \apj, 346, 860

\reference{Ste} Stewart, R. T., Haynes, R. F., Gray, A. D., \& Reich, W. \ 1994, \apjl, 432, L39

\reference{Sub} Subrahmanyan, R., \& Goss, W. M. \ 1995, MNRAS, 275, 755

\reference{Tho} Thompson., C., \& Duncan, R. C. \ 1995, MNRAS, 275, 255

\reference{Tro} Troland, T. H., \& Heiles, C. \ 1982, ApJ, 260, L19

\reference{Wall97} Wallin, B. K., \& Watson, W. D. \ 1997, ApJ, 481, 832

\reference{War99} Wardle, M., Yusef-Zadeh, F. \& Geballe, T. R. \ 1999, in ASP Conf. Series, 
Vol. 186. The Central Parsecs, Falcke, H., Cotera, A., Duschl, W. J., Melia, F. \& Rieke, M. J., 
ed., 432

\reference{Wardle} Wardle, M., \ 1999, ApJ, 527, L109

\reference{Wei} Weiler, K. W., \& Panagia, N. \ 1980, A\&A, 90, 269

%\reference{Whi} Whiteoak, J. B. Z. \& Green, A. J. \ 1996, \aaps, 118, 329

\reference{Woo} Woods, P., Kouveliotou, C., Van Paradijs, J. Hurley, K., Kippen, R. M., 
Finger, M. H., Briggs, M. S., Dieters, S., \& Fishman, G. J. \ 1999, \apj, 519, L139   

\reference{Yus96} Yusef-Zadeh, F., Roberts, D. A., Goss, W. M., Frail, D. A., \& Green, A. J. 
\ 1996, ApJ, 466, L25

\reference{Yus99a} Yusef-Zadeh, F., Roberts, D. A., Goss, W. M., Frail, D. A., \& Green, A. J. 
\ 1999a, ApJ, 512, 230

\reference{Yus99b} Yusef-Zadeh, F., Goss, W. M., Roberts, D. A., Robinson, B. \& Frail, D. A. 
\ 1999b, ApJ, in press

\reference{Zwe} Zweibel, E. G., \& McKee, C. F. \ 1995, \apj, 439, 779


 
\end{references}
\end{document}